\documentclass[
 aps, pra,
 amsmath,amssymb,
 11pt,
 final,
tightenlines,
 twoside,
 twocolumn,
 nofloats,
nofootinbib,
 superscriptaddress,
showkeys,
showkeywords,
 ]
{revtex4-2}
\usepackage{booktabs}
\usepackage{upgreek}
\usepackage{url}
\usepackage{natbib}
\usepackage{appendix}
\usepackage[T1]{fontenc}
\usepackage[utf8x]{inputenc}
\usepackage[english]{babel}
\usepackage{graphicx}
\usepackage{dcolumn}
\usepackage{bm}
\usepackage{longtable}
\input{maik.rty}

\setcitestyle{authoryear,round}
\def\squareforqed{\hbox{\rlap{$\sqcap$}$\sqcup$}}

\def\sq{\ifmmode\squareforqed\else{\unskip\nobreak\hfil
\penalty50\hskip1em\null\nobreak\hfil\squareforqed
\parfillskip=0pt\finalhyphendemerits=0\endgraf}\fi}

\def\utw{\smash{\rlap{\lower5pt\hbox{$\sim$}}}}

\def\udtw{\smash{\rlap{\lower6pt\hbox{$\approx$}}}}

\def\diameter{{\ifmmode\mathchoice
{\ooalign{\hfil\hbox{$\displaystyle/$}\hfil\crcr
{\hbox{$\displaystyle\mathchar"20D$}}}}
{\ooalign{\hfil\hbox{$\textstyle/$}\hfil\crcr
{\hbox{$\textstyle\mathchar"20D$}}}}
{\ooalign{\hfil\hbox{$\scriptstyle/$}\hfil\crcr
{\hbox{$\scriptstyle\mathchar"20D$}}}}
{\ooalign{\hfil\hbox{$\scriptscriptstyle/$}\hfil\crcr
{\hbox{$\scriptscriptstyle\mathchar"20D$}}}}
\else{\ooalign{\hfil/\hfil\crcr\mathhexbox20D}}%
\fi}}

\newcommand{\aap}{Astron. and Astrophys. }

\newcommand{\aj}{Astron.~J. }

\newcommand{\mnras}{Monthly Notices Royal Astron. Soc. }

\newcommand{\pasp}{Publ. Astron. Soc. Pacific }

\begin{document}

\selectlanguage{english}



\title{Discovery of new $\delta$\, Scuti Stars}

\author{\firstname{\"{O}.}~\surname{K{\i}rm{\i}z{\i}ta\textcommabelow{s}}}
 \email{ozlemkirmizitas1907@gmail.com}
 \affiliation{\c{C}anakkale Onsekiz Mart University, Faculty of Sciences and Arts, Physics Department, 17100, Canakkale, Turkey}

\author{\firstname{S.}~\surname{\c{C}avu\textcommabelow{s}}}
 \affiliation{\c{C}anakkale Onsekiz Mart University, Faculty of Sciences and Arts, Physics Department, 17100, Canakkale, Turkey}

\author{\firstname{F.}~\surname{Kahraman Ali\c{c}avu\c{s}}}
\affiliation{\c{C}anakkale Onsekiz Mart University, Faculty of Sciences and Arts, Physics Department, 17100, \c{C}anakkale, Turkey}
 \affiliation{\c{C}anakkale Onsekiz Mart University, Astrophysics Research Center and Ulup{\i}nar Observatory, TR-17100, \c{C}anakkale, Turkey}

\begin{abstract}
 Pulsating stars are remarkable objects for stellar astrophysics. Their pulsation frequencies allow us to probe the internal structure of stars. One of the most known groups of pulsating stars is $\delta$\,Scuti variables which could be used to understand the energy transfer mechanism in A-F type stars. Therefore, in the current study, we focused on the discovery of $\delta$\,Scuti stars. For this investigation, we followed some criteria. First, we inspected TESS database by eye and discovered some single stars that exhibit pulsation like behaviour. Our second criterion is $T_{\rm eff}$ and $\log g$ range. The $\delta$\,Scuti stars generally have $T_{\rm eff}$ and $\log g$ value in a range of 6300\,$-$\,8500\,K and 3.2\,$-$\,4.3, respectively. Hence, we selected the stars which have TIC $T_{\rm eff}$ and $\log g$ values in these ranges. The other criterion is the pulsating frequency. A frequency analysis was performed for all  the candidate stars. In addition, $M_{V}$, $L$ and also $M_{bol}$ parameters of the target stars were determined to calculate the pulsation constants and show their positions in the H-R diagram. The final pulsation type classification was made by considering the frequency ranges and pulsation constants of the stars. As a result of the study, five $\delta$\,Scuti, one $\gamma$\,Doradus and four hybrid systems were discovered.   
\end{abstract}

\maketitle 

\section{INTRODUCTION}

Asteroseismology is a great tool to investigate interior structure of stars via their oscillation modes. The main targets of asteroseismology is the pulsating variables. There are many pulsating stars known for decades such as $\beta$\,Cephei, $\delta$\,Scuti and Cepheid stars. Inside the group of pulsating variables, the main sequence A-F type oscillating stars occupy a significant place in asteroseismology, as they are located on transition region where the convection envelopes of stars turn into radiative envelope \citep{2010aste.book.....A}. To understand the mechanism operating in that region, investigations of these A-F type pulsating stars become very important. 

Mainly, there are two main-sequence A-F type pulsating variables; $\delta$\,Scuti ($\delta$\,Sct) and $\gamma$\,Doradus ($\gamma$\,Dor) stars. The $\delta$\,Sct stars have a spectral type of A0-F5 with luminosity class changing from dwarf to giant \citep{2013AJ....145..132C}. These variables exhibit radial and non-radial oscillations with a general frequency range of $5-50$\,d$^{-1}$ \citep{2000ASPC..210....3B, 2010aste.book.....A}. The $\gamma$\,Dor stars are dwarf and/or sub-dwarf variables with a spectral type of A7-F5 \citep{1999PASP..111..840K}. These variables exhibit non-radial pulsations with a frequency generally lower than 5\,d$^{-1}$ \citep{1999PASP..111..840K,2011A&A...534A.125U}. The instability strips of these A-F type pulsating stars are placed to the lower part of the classical instability strip and they are partially overlap. In this overlapping part, existence of $\delta$\,Sct and $\gamma$\,Dor hybrid stars was proposed \citep{1996A&A...313..851B, 2002MNRAS.333..251H, 2004A&A...414L..17D}. These hybrid variables may also be called A-F type hybrids. These stars show both $\delta$\,Sct and $\gamma$\,Dor type oscillations at the same time. 

A great revolution on the study of A-F type pulsating variables have been achieved by the contribution of space telescopes. Especially, studies of high-quality {\sl Kepler} \citep{2010Sci...327..977B} data had a significant contribution. \cite{2011A&A...534A.125U} were examined {\sl Kepler} data of 750 A-F type pulsating stars and revealed a general information about these variables and also show that there are some unexpected situations about these pulsators. For example, they pointed out that some $\delta$\,Sct, $\gamma$\,Dor and/or hybrid stars may place outside of theoretically suggested area. This situation is not expected according to the theoretical studies \citep{2004A&A...414L..17D, 2005A&A...435..927D}. 

After {\sl Kepler}, Transiting Exoplanet Survey Satellite (TESS, \citeauthor{2015JATIS...1a4003R}\citeyear{2015JATIS...1a4003R}) started to provide high-quality space data of many stellar systems. The first study of A-F type pulsating stars with TESS data was presented by \cite{2019MNRAS.490.4040A} and properties of these variables were examined. As the TESS almost observed the entire sky, it is a significant tool to investigate these kinds of variables and discover new candidates. Therefore, in this study we present one part of our TESS field search to reveal mainly new $\delta$\,Sct variables. The paper is organized as follows. Information about used TESS data and the criteria of target selection are given in Sect.\,\ref{data}. The time-series analysis is present in Sect.\,\ref{frekans}. Discussion and conclusions are introduced in Sect.\,\ref{conc}.

\begin{table*}
\begin{center}
\caption[]{Information about the targets. The parameters and their average errors were taken from TIC \citep{2019AJ....158..138S}.}\label{first}
 \begin{tabular}{llcccccr}
   \midrule
  \hline\noalign{\smallskip}
 TIC & Other & RA      & Dec(J2000) & TESS    &   $T_{\rm eff}$ & $\log g$ & TESS     \\
     & name  &  (J2000)& (J2000)    & (mag\,$\pm$\,0.006)   &  (K\,$\pm$\,136)  & \,$\pm$\,0.08          & Sectors \\
  \hline\noalign{\smallskip}
25537276  &HD 221009      & 23$^{h}$ 28$^{m}$ 08$^{s}$.7 &+49$^{o}$ 02$^{'}$ 54$^{''}$.8 & 8.52 & 6866 & 3.36 & 16 \\
177422294 &HD 223901      & 23$^{h}$ 53$^{m}$ 29$^{s}$.7 &+43$^{o}$ 37$^{'}$ 57$^{''}$.0 & 8.48 & 7223 & 3.34 & 17 \\
252554307 &TYC 3626-505-1 & 23$^{h}$ 00$^{m}$ 27$^{s}$.0 &+48$^{o}$ 24$^{'}$ 57$^{''}$.2 & 8.46 & 6812 & 3.81 & 16 \\
279874050 &HD 222170      &	23$^{h}$ 37$^{m}$ 48$^{s}$.7 &+67$^{o}$ 55$^{'}$ 39$^{''}$.1 & 8.20 & 6873 &      & 17, 18, 24, 25 \\
308447073 &HD 9469        & 01$^{h}$ 34$^{m}$ 01$^{s}$.9 &+48$^{o}$ 34$^{'}$ 45$^{''}$.1 & 8.74 & 7539 & 4.29 & 18 \\
367910480 &HD 222386      & 23$^{h}$ 39$^{m}$ 10$^{s}$.1 &+75$^{o}$ 17$^{'}$ 34$^{''}$.3 & 5.87 & 8476 & 4.03 & 18, 19, 24, 25 \\
370599803 &HD 10941       & 01$^{h}$ 48$^{m}$ 55$^{s}$.2 &+54$^{o}$ 56$^{'}$ 53$^{''}$.3 & 8.09 & 7394 & 3.76 & 18 \\
395520454 &HD 10880       & 01$^{h}$ 50$^{m}$ 11$^{s}$.0 &+73$^{o}$ 07$^{'}$ 08$^{''}$.2 & 8.88 & 7353 &      & 25 \\
400502366 &HD10085        & 01$^{h}$ 39$^{m}$ 55$^{s}$.2 &+55$^{o}$ 38$^{'}$ 16$^{''}$.0 & 8.15 & 8261 & 3.54 & 18 \\
431375592 &HD 219429      &	23$^{h}$ 15$^{m}$ 11$^{s}$.3&+48$^{o}$  13$^{'}$ 05$^{''}$.9 & 8.72 & 6411 & 4.04 & 16, 17 \\
  \noalign{\smallskip}\hline
\end{tabular}
\label{tablo1}
\end{center}  
\end{table*}

\section{Data and Target Selection}
\label{data}

In this study, TESS data were used for discovering new $\delta$\,Sct stars. The main mission of TESS is detecting new exoplanets orbiting around close stars \citep{2015JATIS...1a4003R}. TESS was launched from Cape Canaveral Base in the USA on April 2018, with the Falcon 9 rocket produced by SpaceX company. TESS has the same working logic with the \textit{Kepler} space telescope, however, it scans a wider area. TESS satellite has four identical CCD cameras and each has a 24$\times$24 degree field of view \citep{2015JATIS...1a4003R}. Both northern and southern hemispheres have been observed by TESS and it has provided many data of stellar systems in addition to revealing new exoplanets \citep[e.g.][]{2019MNRAS.490.4040A}. TESS photometric data has been obtained in 2-min short (SC) and 30-mins long (LC) cadences during first two years mission. Those data are released in the Barbara A. Mikulski Archive for Telescopes (MAST)\footnote{https://mast.stsci.edu} archive. 

In the current study, we present a part of our northern TESS filed research to discover unknown $\delta$\,Sct type variables. We took the TESS data from MAST archive. Only SC TESS data were checked to find these kinds of variable stars. First criterion in our target selection is determining some stars showing pulsating like variation in their light curve. In this study we only focused on single pulsating variables and eliminated the eclipsing binaries. The second criterion is the effective temperature ($T_{\rm eff}$) and also surface gravity ($\log g$) range. The $\delta$\,Sct stars have generally $T_{\rm eff}$ and $\log g$ in the range of 6300\,-\,8500\,K and 3.2\,-\,4.3 \citep{2001A&A...366..178R}, so  we controlled these values of the systems found in the first target selection. The $T_{\rm eff}$ and $\log g$ values of the selected systems were taken from the TESS Input Catalog \citep[TIC,][]{2019AJ....158..138S}. The stars having $T_{\rm eff}$ and $\log g$ values in the given ranges were chosen as a result of this target selection criterion. The final criterion is determining the unknown systems in the literature. Therefore as a final step we checked the literature information of the stars found in the previous selection parts and determined the unknown systems. The final target list was obtained by considering these three criteria. This list is given in Table\,\ref{first}.

To confirm the pulsational variability of the selected systems one needs to analyse TESS data in detail. Therefore, all SC TESS data of the systems were collected. The the simple aperture photometry (SAP) fluxes were taken into account and these data were converted into magnitude. The final TESS data were also normalised by using a polynomial fit to eliminate possible instrumental effects.

\section{Frequency Analysis} \label{frekans}

To reveal the pulsation characteristic of the selected targets a frequency analysis was performed. In the frequency analysis, all available SC data of the systems were used because the Nyquist frequency of SC data reach to $\sim$360\,d$^{-1}$. Considering the pulsation period range of the $\delta$\,Sct stars, the SC data are suitable to searching for $\delta$\,Sct-type oscillations. In the frequency analysis, the {\sc Period04} program \citep{2005CoAst.146...53L} was used. Basing on the Fourier and non-linear least square fitting the program searches for single and multiple frequencies with the "pre-whitening" method \citep{2009MNRAS.398.1339H}. The program is also convenient to detect possible harmonic and combination frequencies. 
According to the criterion given by \cite{1993A&A...271..482B}, the frequencies having a signal-to-noise (S/N) ratio over 4 is considered as significant. However, in a recent study of \cite{2021AcA....71..113B}, it was showed that for the space-based TESS data S/N ratio should be higher than typically used Breger criterion to detect significant frequencies. Therefore, taking into account the results of this study we considered the frequencies as significant if they have S/N ratio over 4.5.

\begin{figure*}
 \centering
 \begin{minipage}[b]{0.24\textwidth}
  \includegraphics[height=3.5cm, width=1\textwidth]{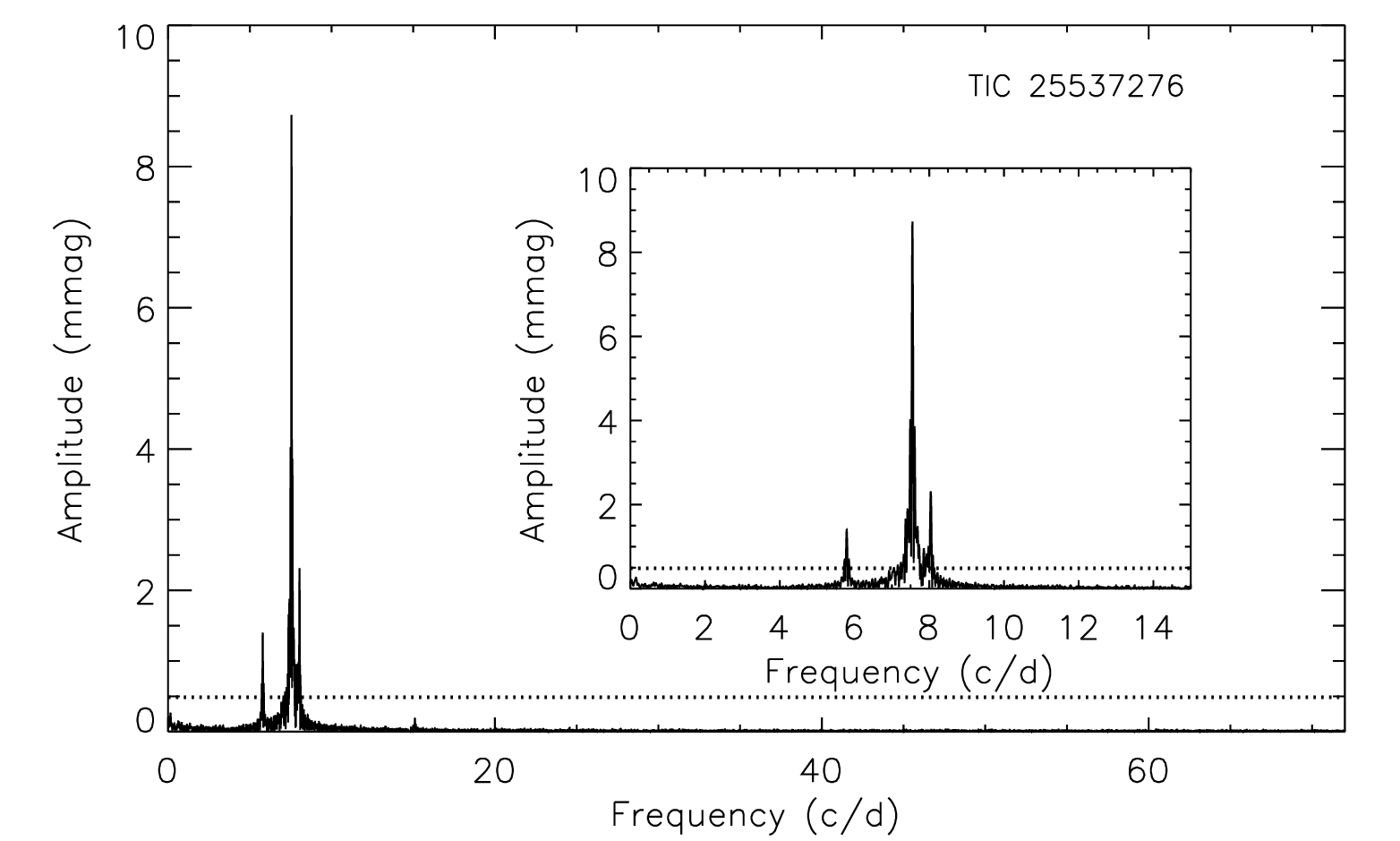}
 \end{minipage}
  \begin{minipage}[b]{0.24\textwidth}
  \includegraphics[height=3.5cm, width=1\textwidth]{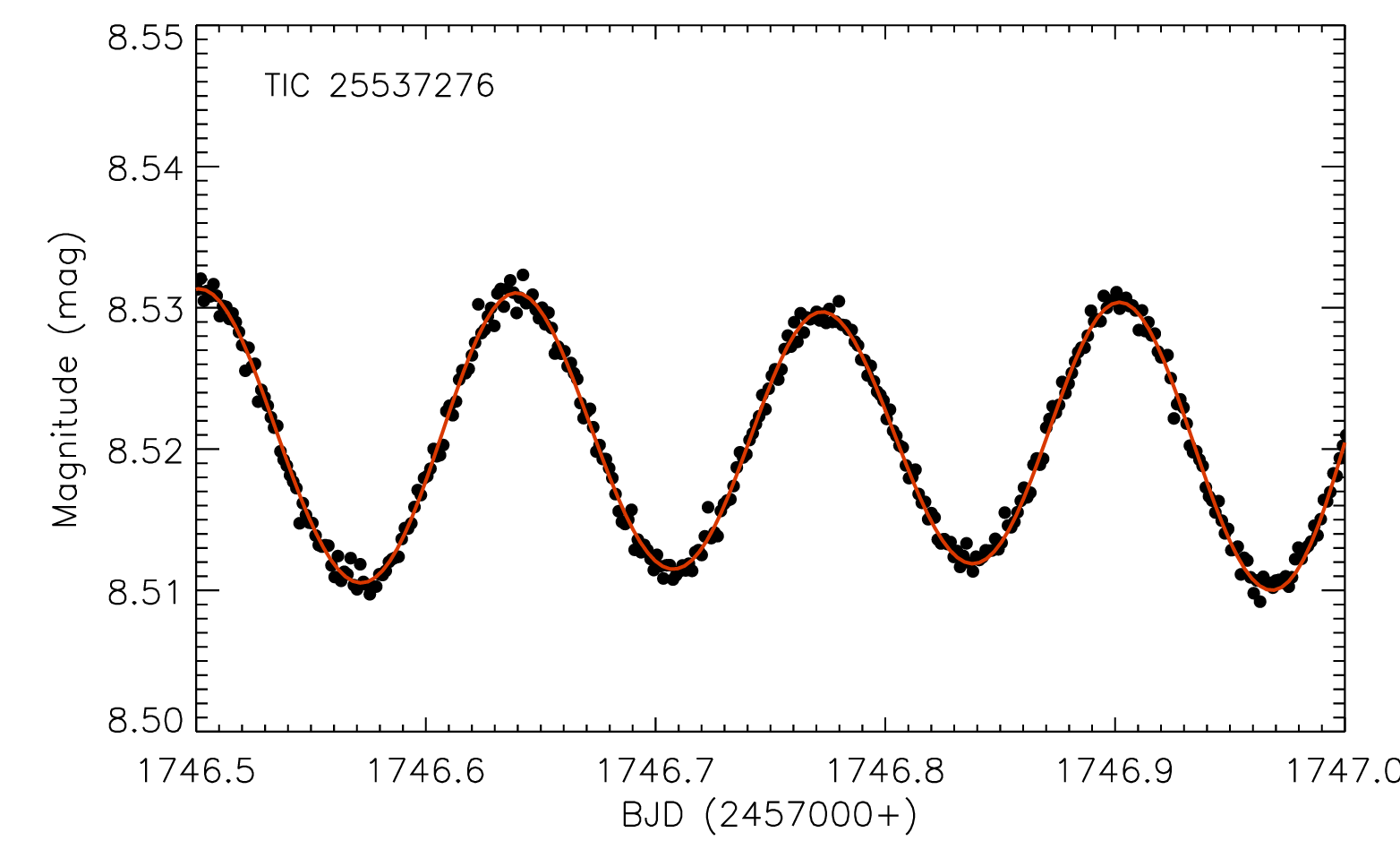}
 \end{minipage}
 \begin{minipage}[b]{0.24\textwidth}
  \includegraphics[height=3.5cm, width=1\textwidth]{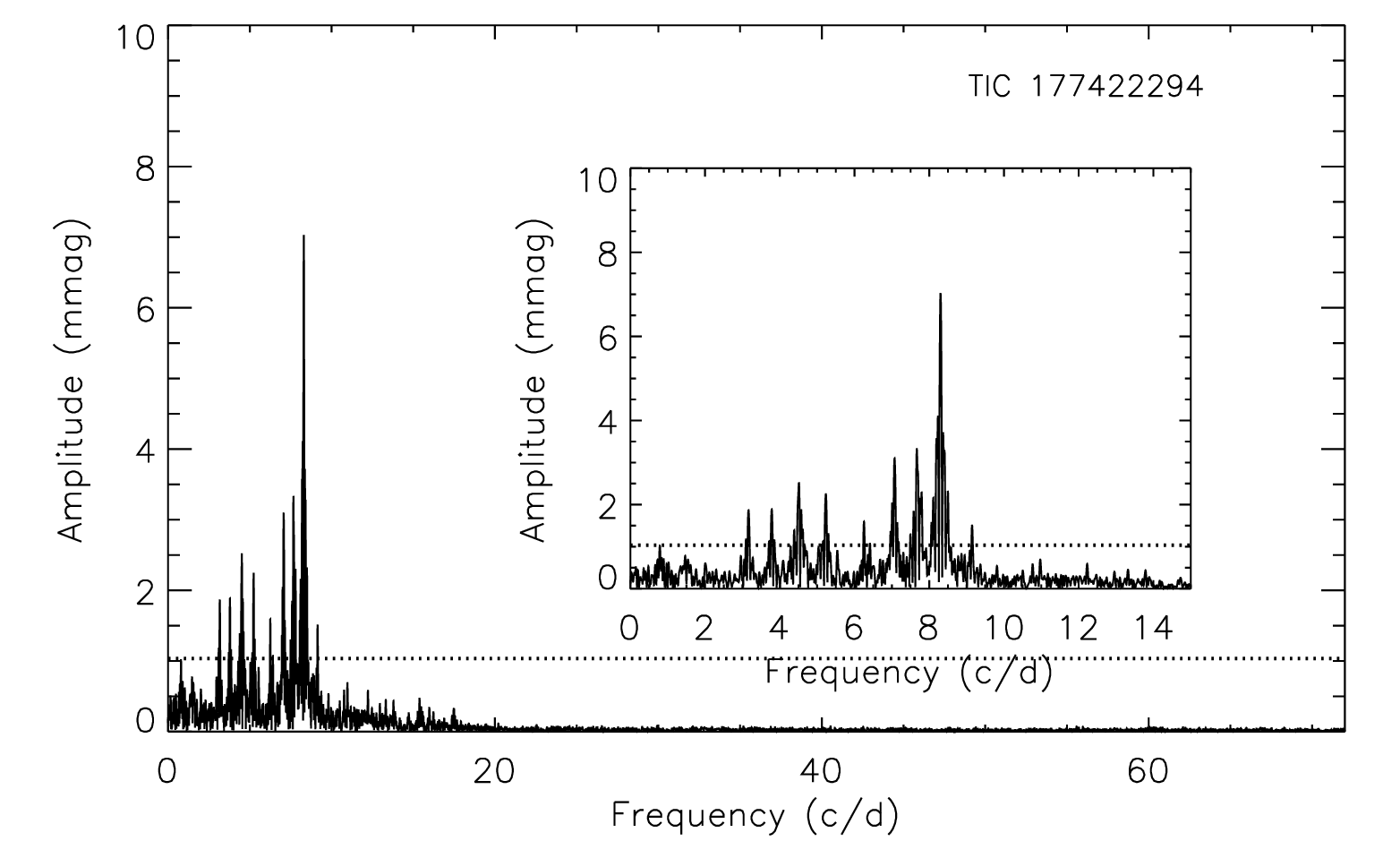}
  \end{minipage}
  \begin{minipage}[b]{0.24\textwidth}
  \includegraphics[height=3.5cm, width=1\textwidth]{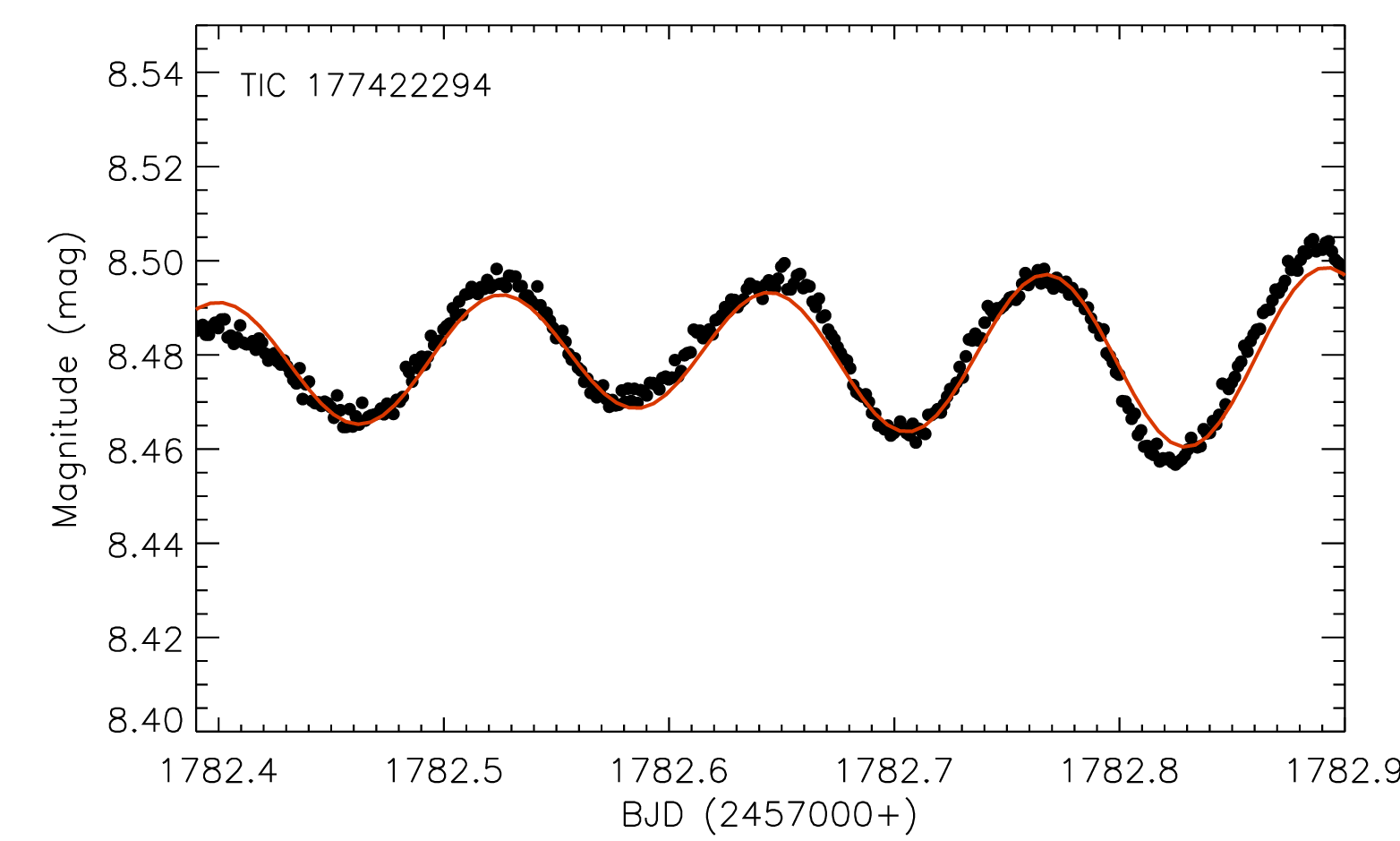}
  \end{minipage}
 \begin{minipage}[b]{0.24\textwidth}
  \includegraphics[height=3.5cm, width=1\textwidth]{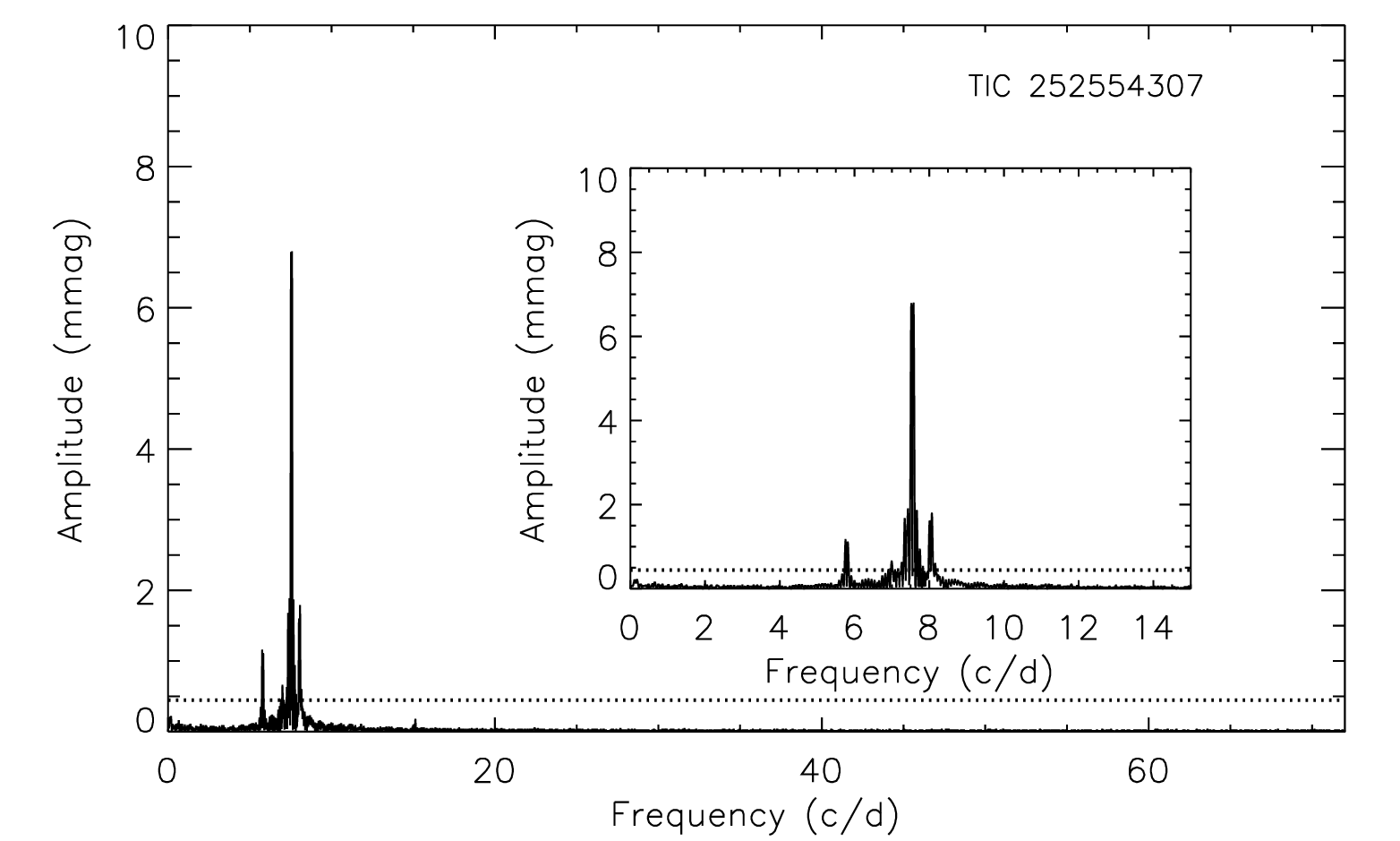}
 \end{minipage}
 \begin{minipage}[b]{0.24\textwidth}
  \includegraphics[height=3.5cm, width=1\textwidth]{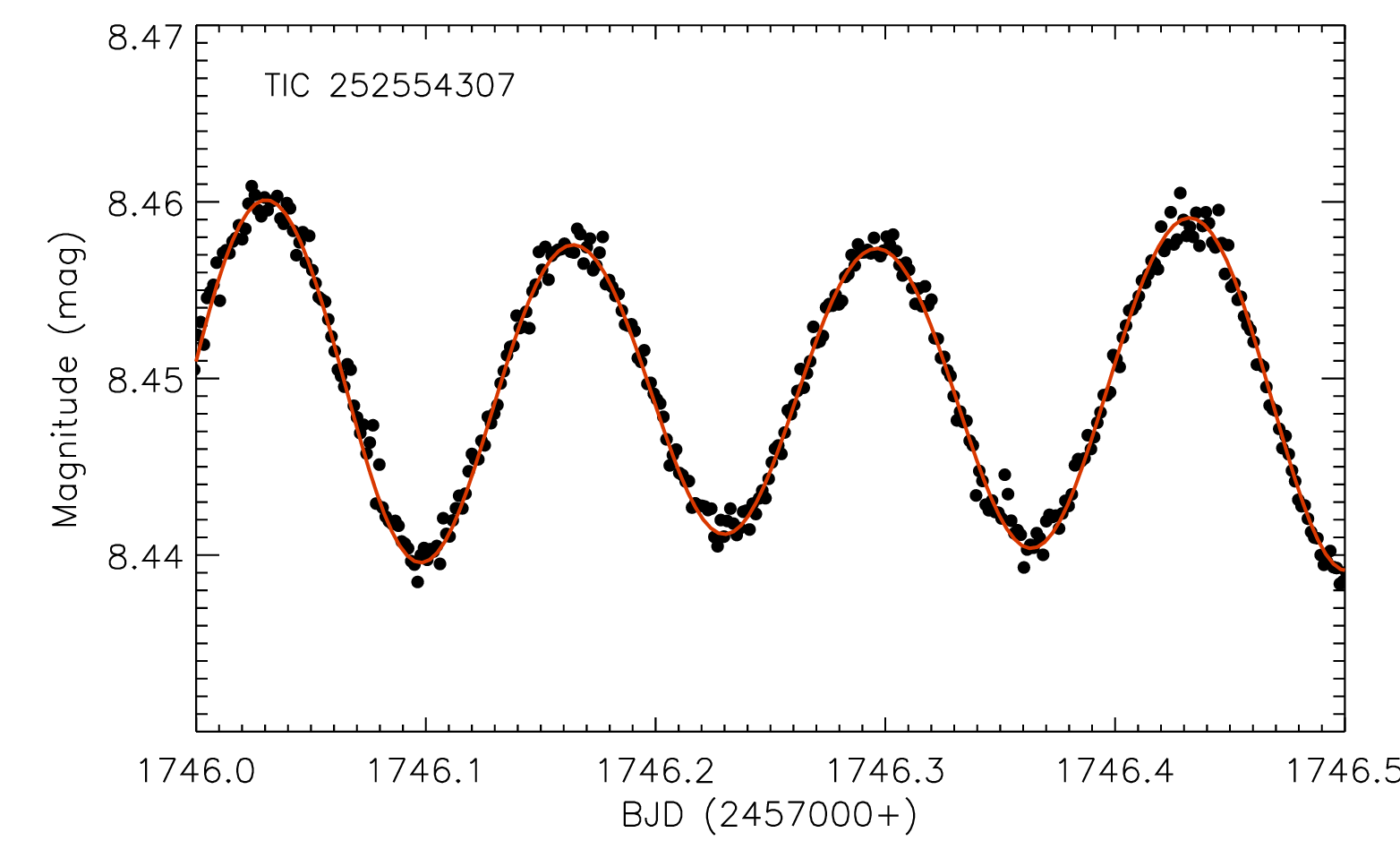}
 \end{minipage}
   \begin{minipage}[b]{0.24\textwidth}
  \includegraphics[height=3.5cm, width=1\textwidth]{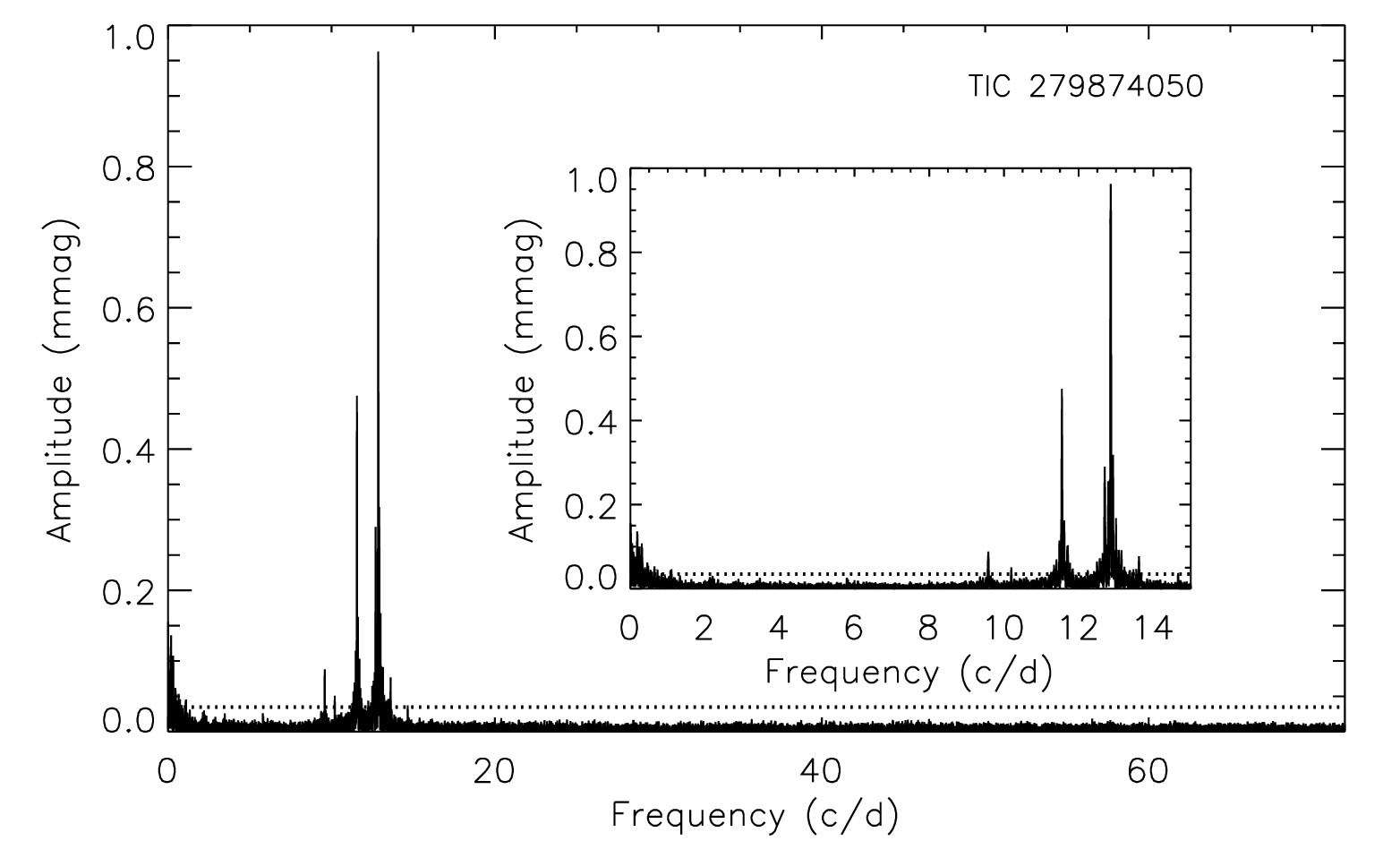}
 \end{minipage}
    \begin{minipage}[b]{0.24\textwidth}
  \includegraphics[height=3.5cm, width=1\textwidth]{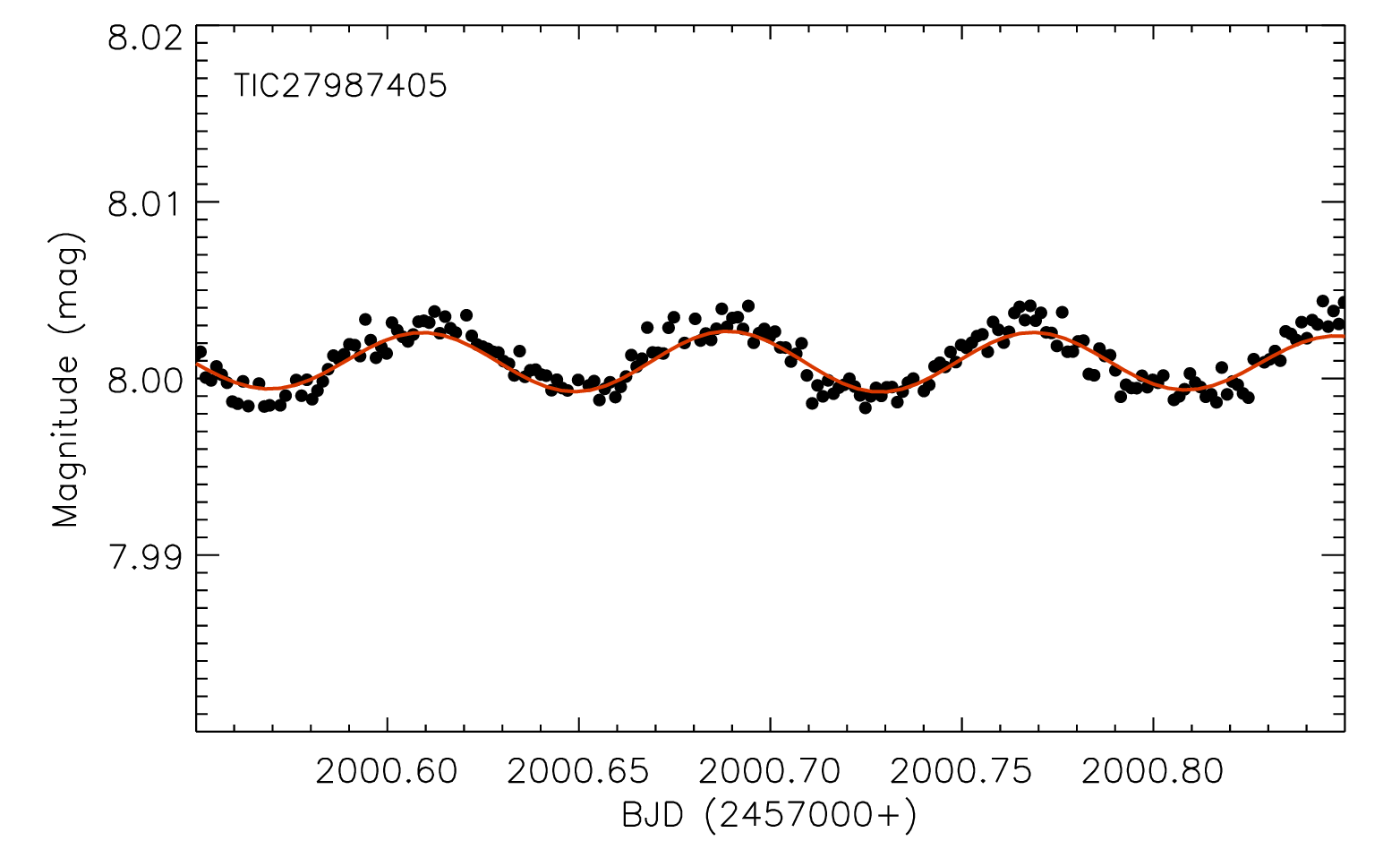}
 \end{minipage}
  \begin{minipage}[b]{0.24\textwidth}
  \includegraphics[height=3.5cm, width=1\textwidth]{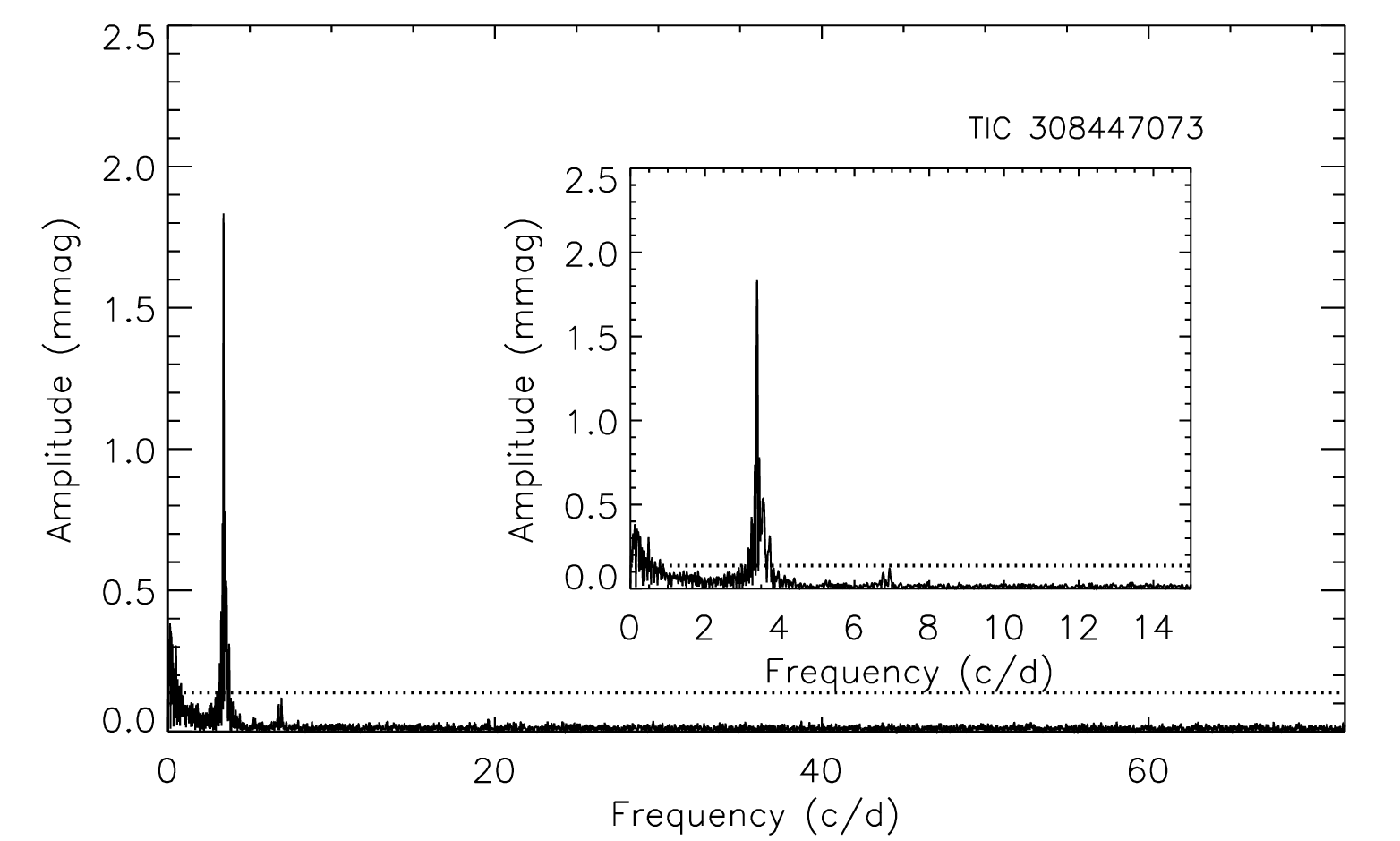}
 \end{minipage}
   \begin{minipage}[b]{0.24\textwidth}
  \includegraphics[height=3.5cm, width=1\textwidth]{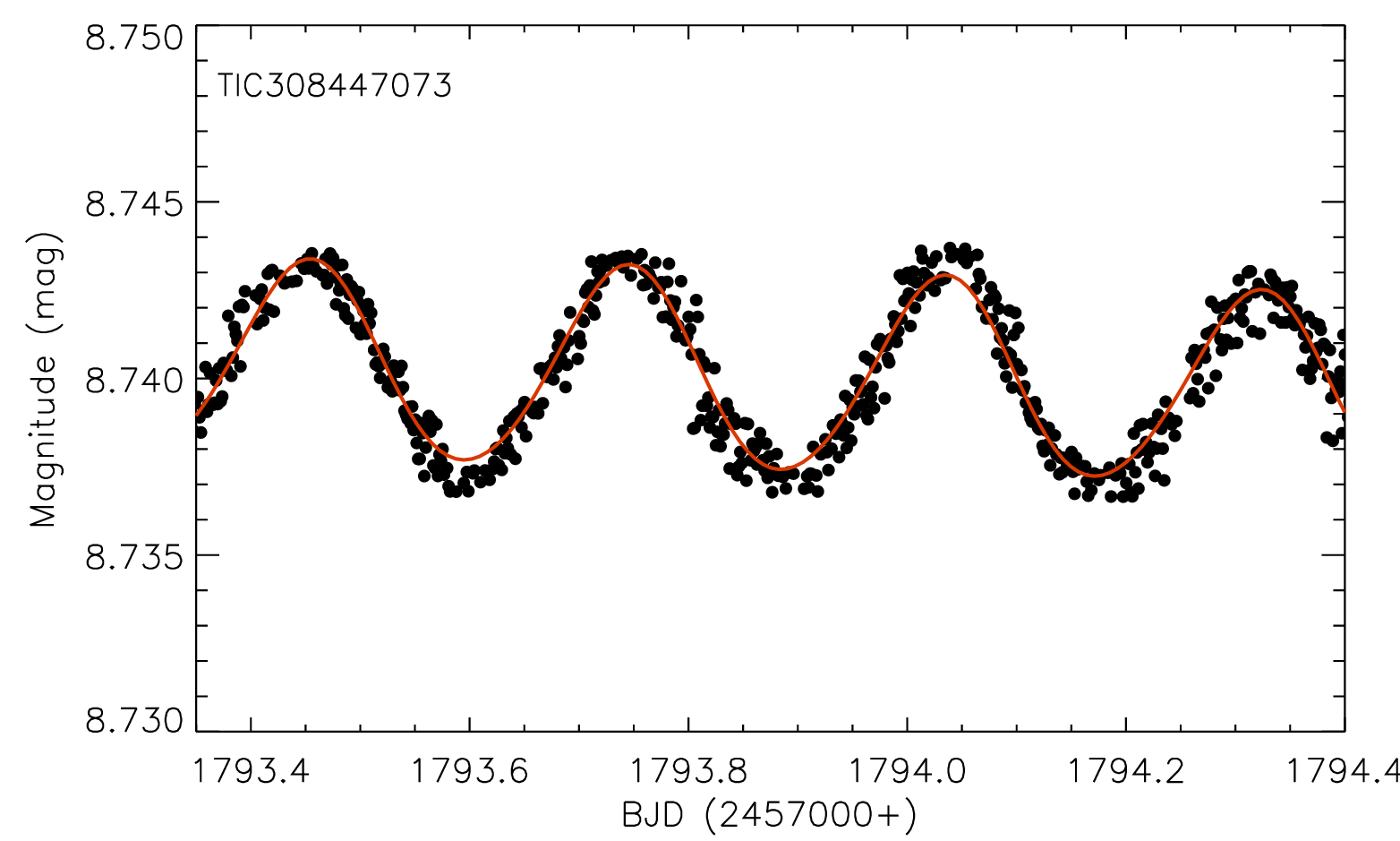}
 \end{minipage}
  \begin{minipage}[b]{0.24\textwidth}
  \includegraphics[height=3.5cm, width=1\textwidth]{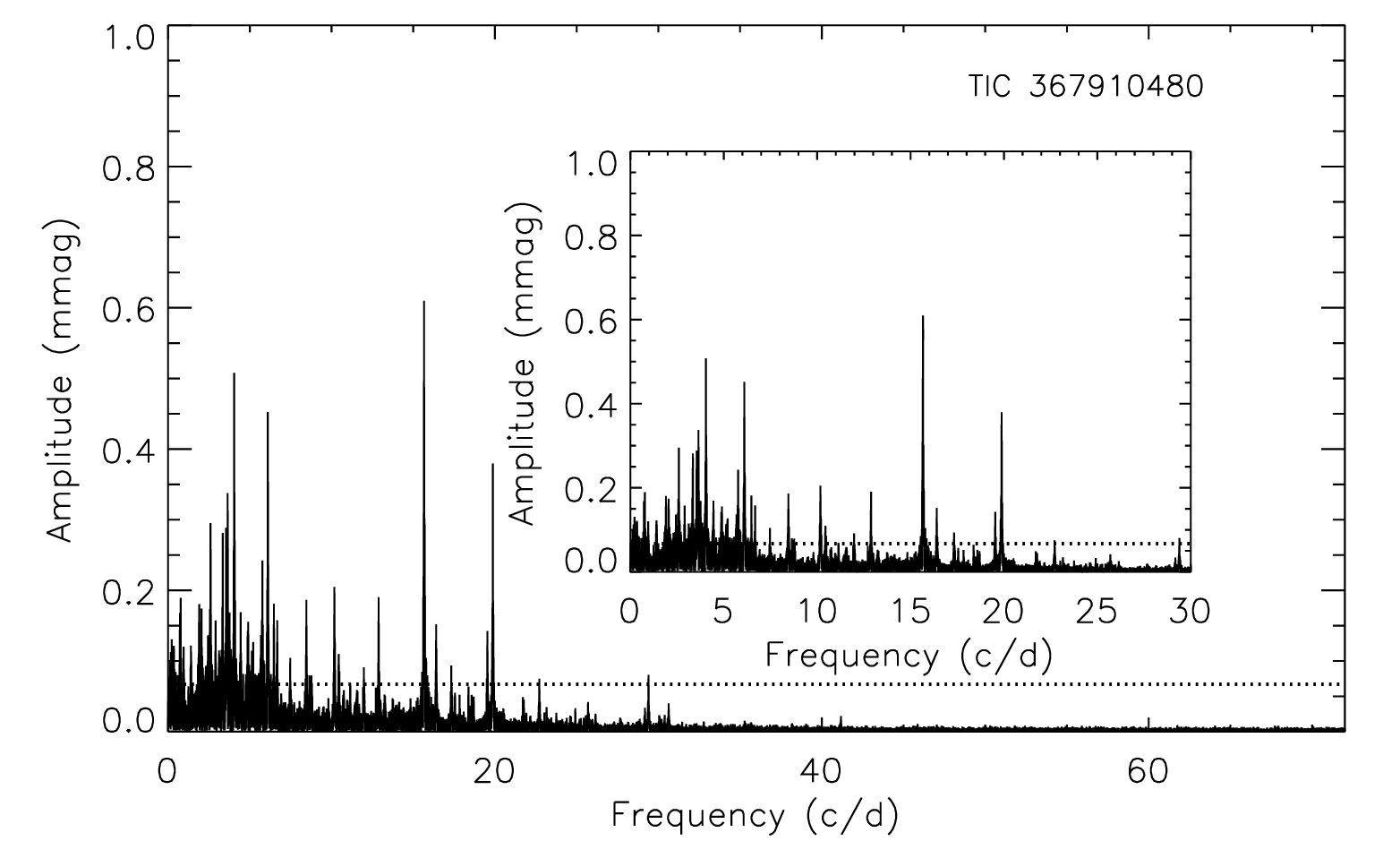}
  \end{minipage}
    \begin{minipage}[b]{0.24\textwidth}
  \includegraphics[height=3.5cm, width=1\textwidth]{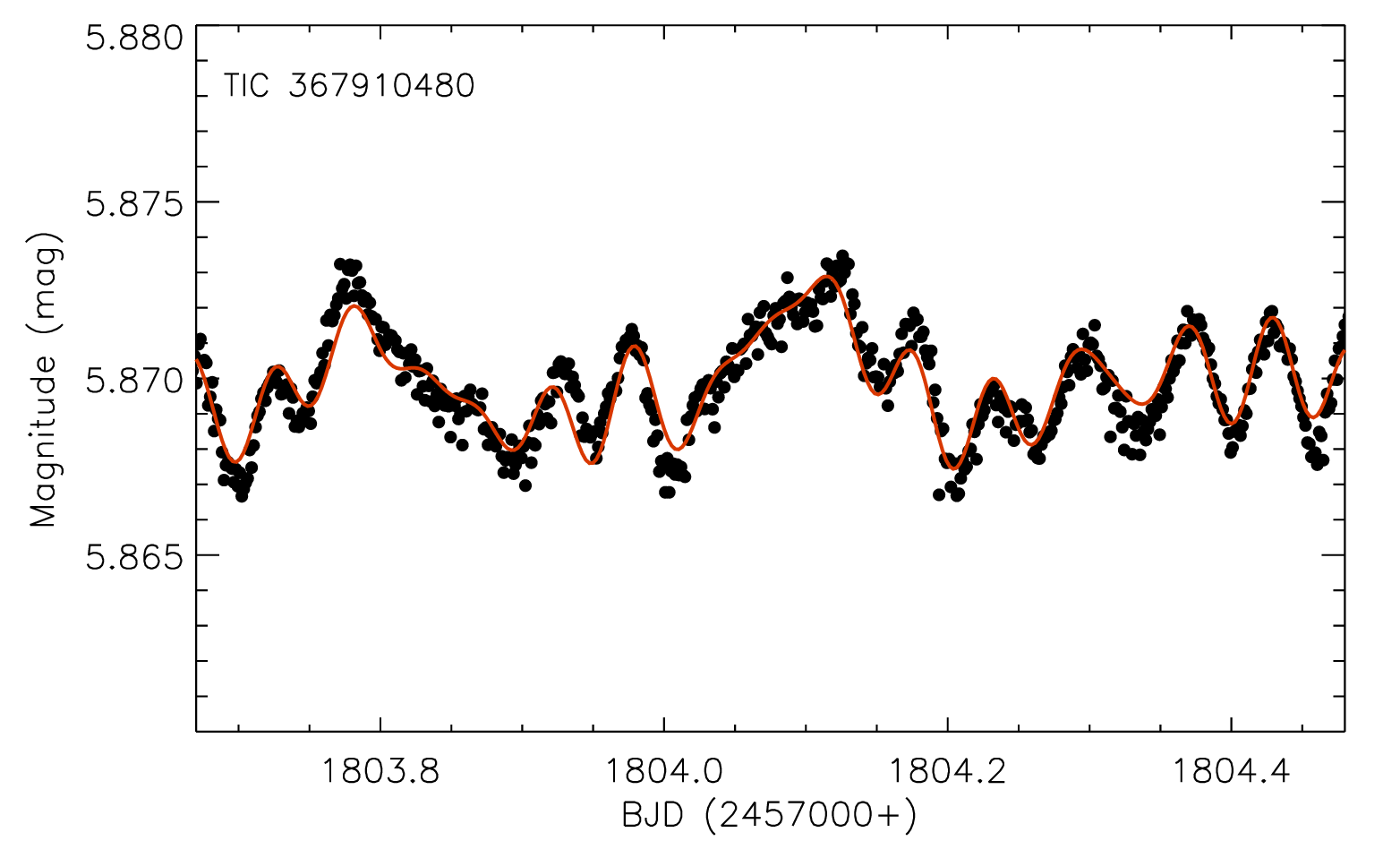}
  \end{minipage}
  \begin{minipage}[b]{0.24\textwidth}
  \includegraphics[height=3.5cm, width=1\textwidth]{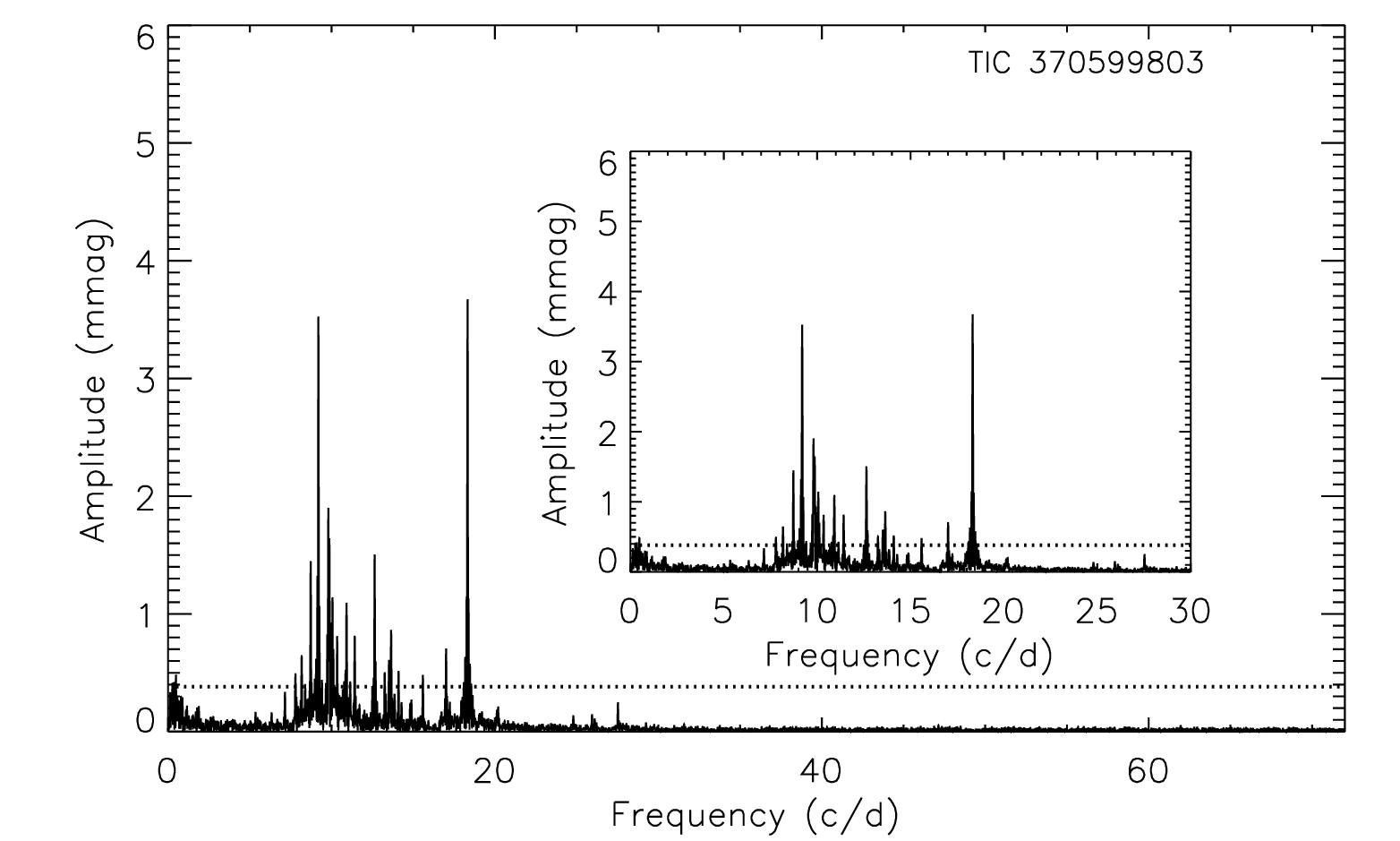}
 \end{minipage}
   \begin{minipage}[b]{0.24\textwidth}
  \includegraphics[height=3.5cm, width=1\textwidth]{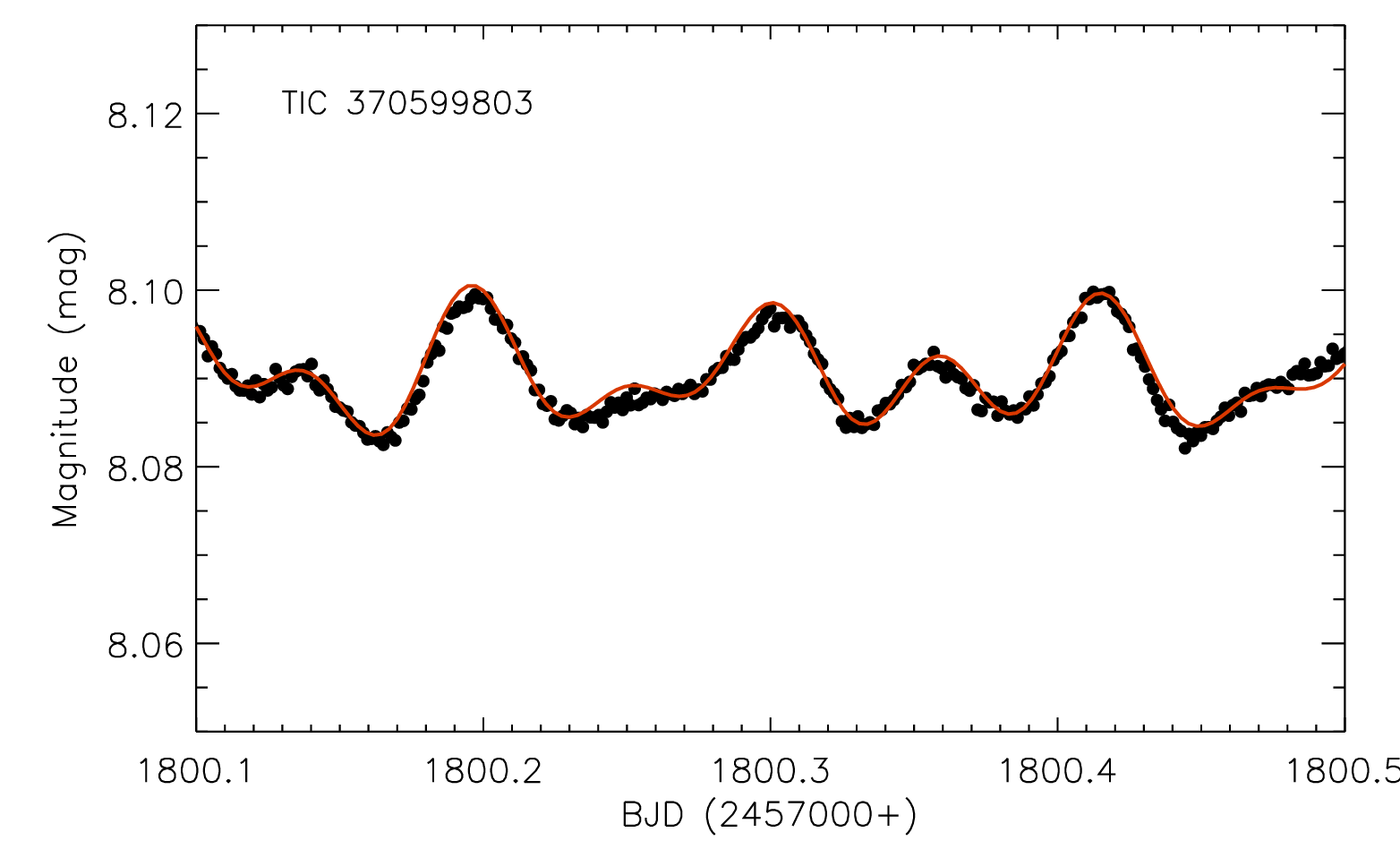}
 \end{minipage}
  \begin{minipage}[b]{0.24\textwidth}
  \includegraphics[height=3.5cm, width=1\textwidth]{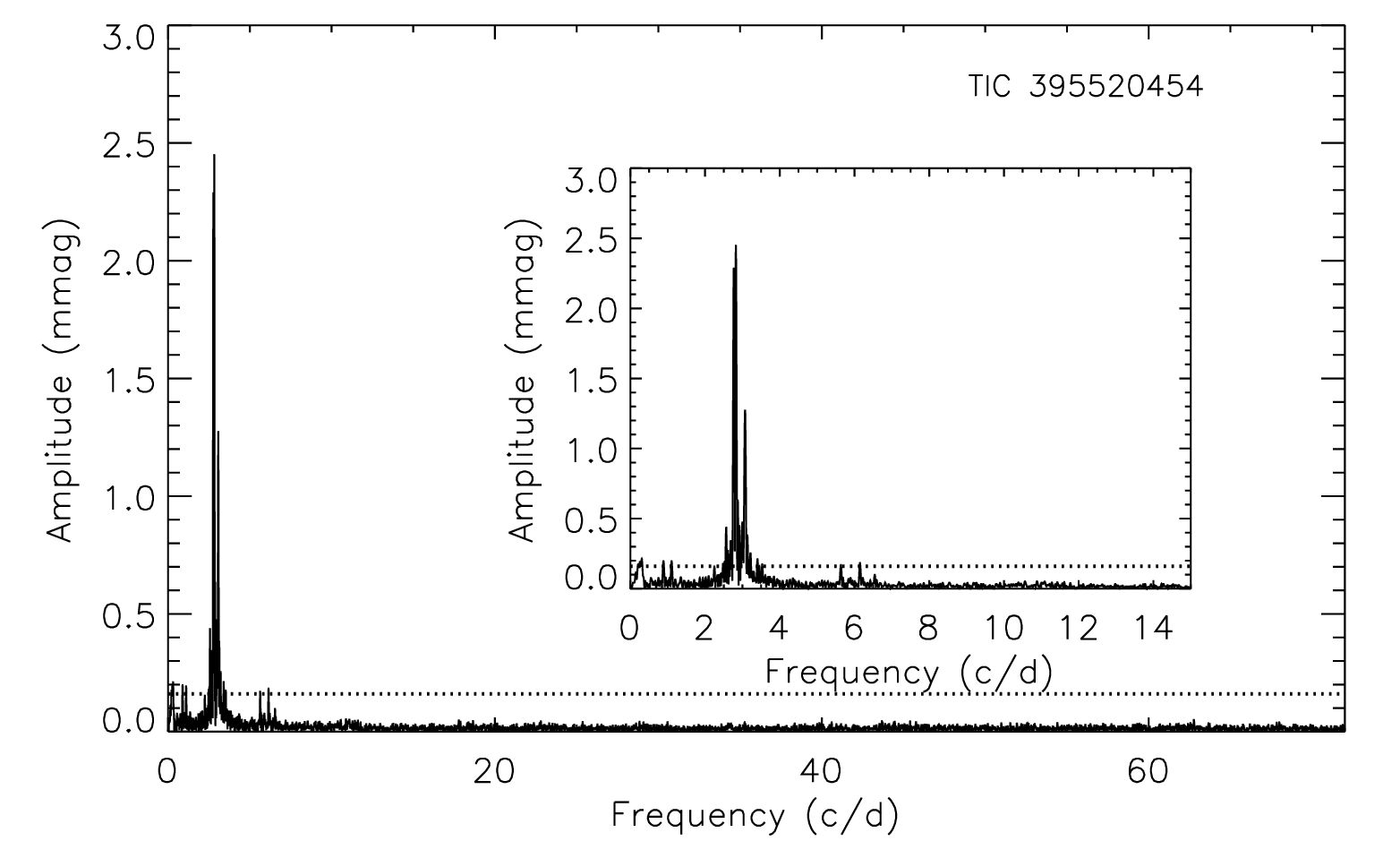}
 \end{minipage}
   \begin{minipage}[b]{0.24\textwidth}
  \includegraphics[height=3.5cm, width=1\textwidth]{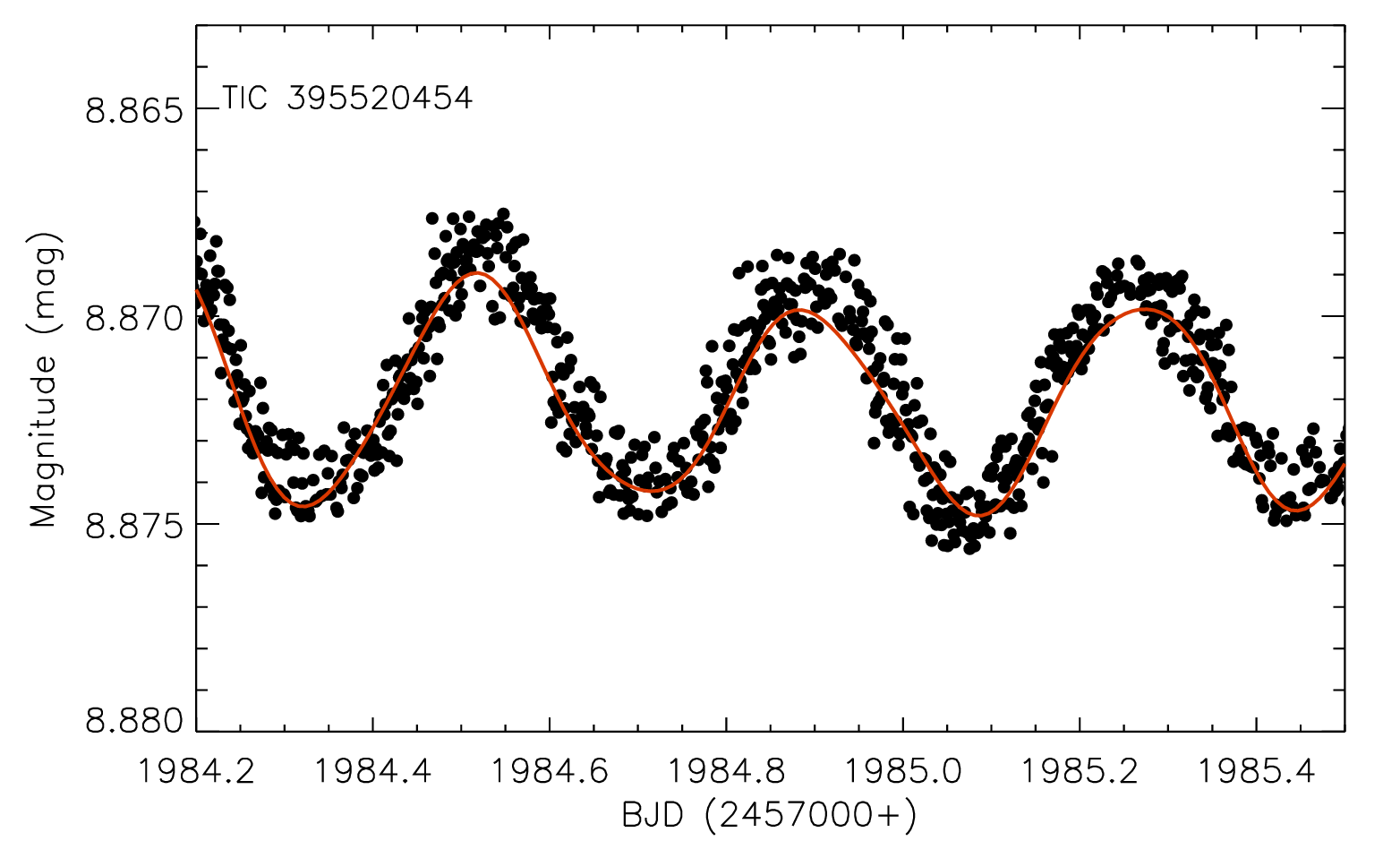}
 \end{minipage}
  \begin{minipage}[b]{0.24\textwidth}
  \includegraphics[height=3.5cm, width=1\textwidth]{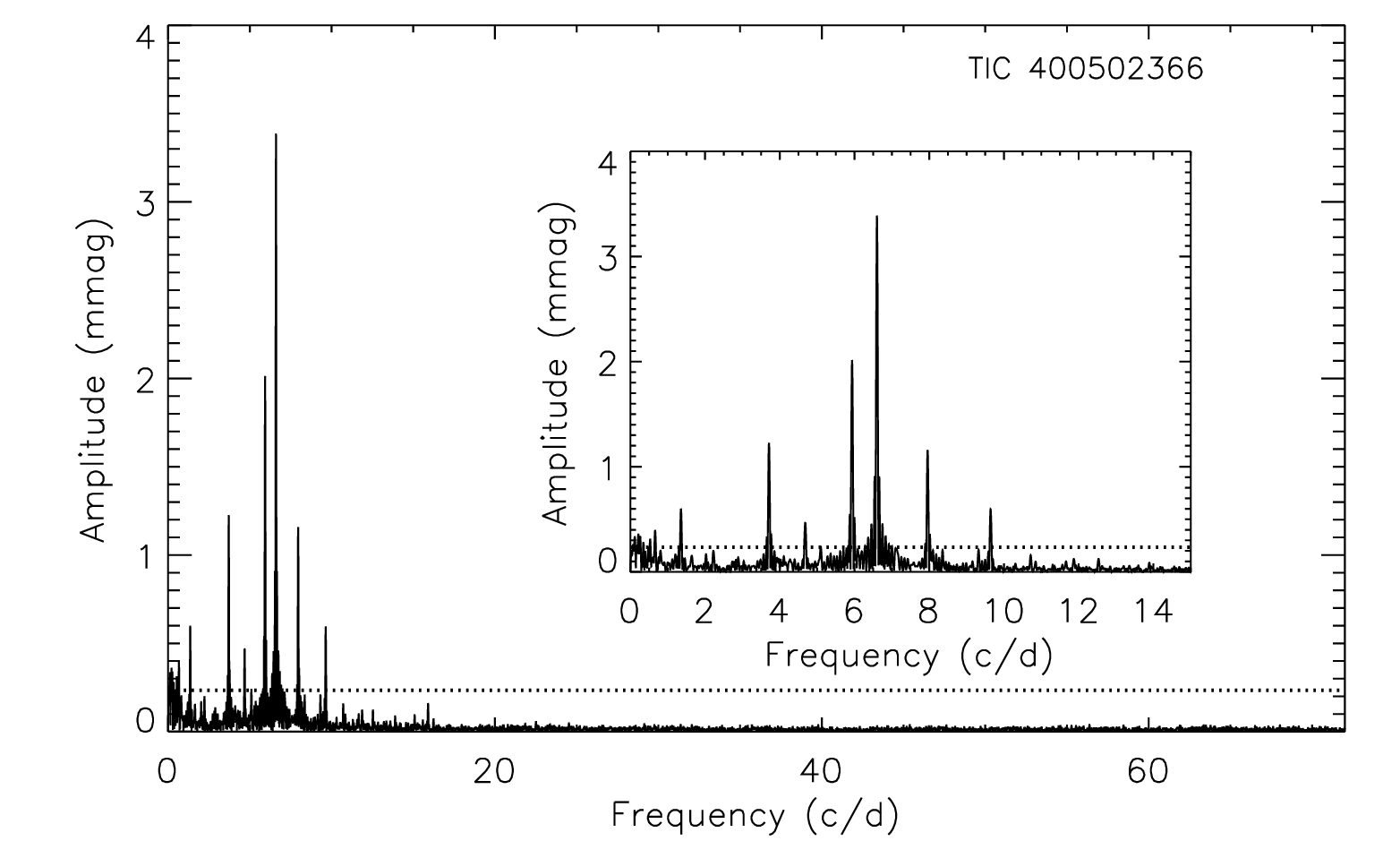}
 \end{minipage}
   \begin{minipage}[b]{0.24\textwidth}
  \includegraphics[height=3.5cm, width=1\textwidth]{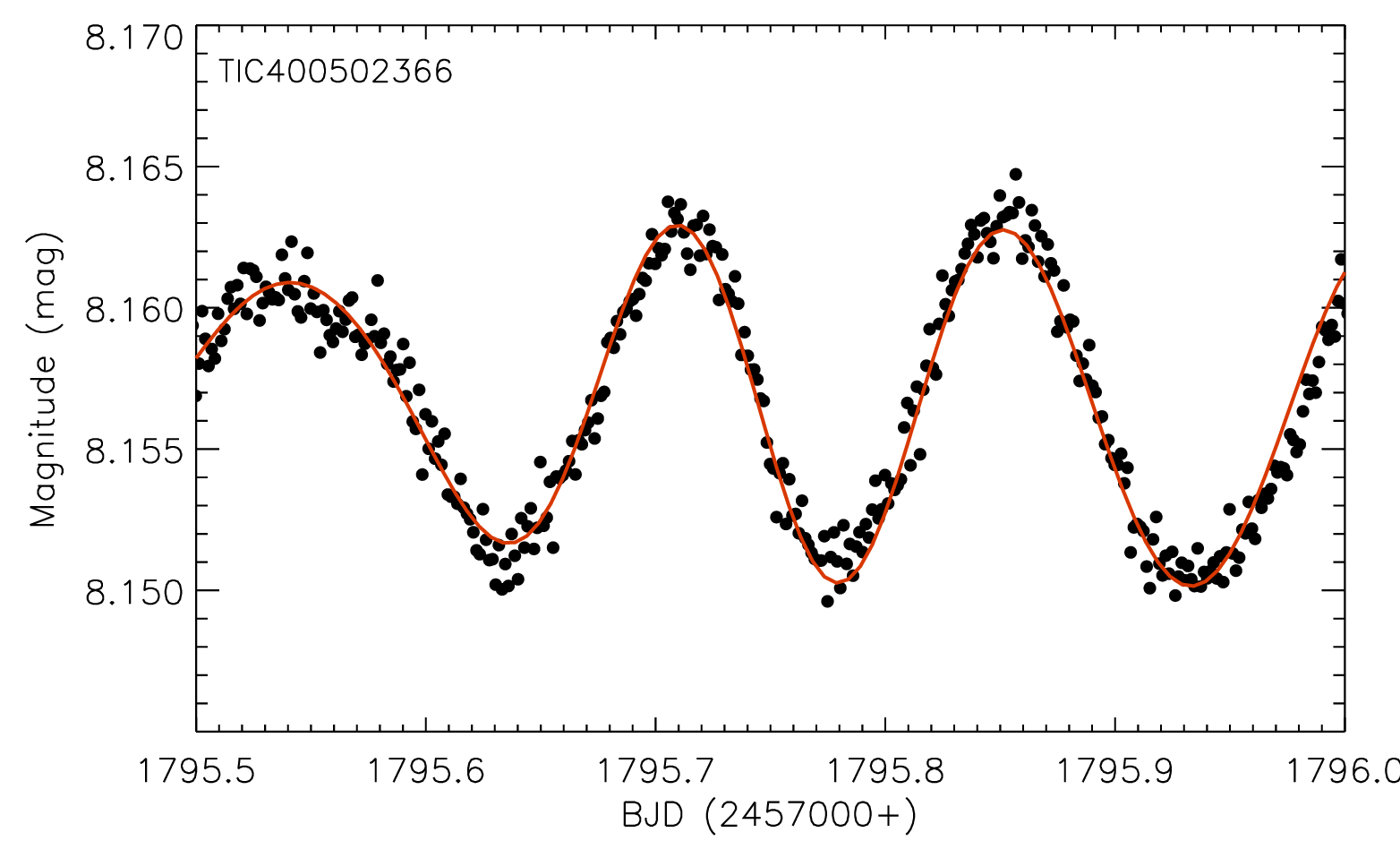}
 \end{minipage}
  \begin{minipage}[b]{0.24\textwidth}
  \includegraphics[height=3.5cm, width=1\textwidth]{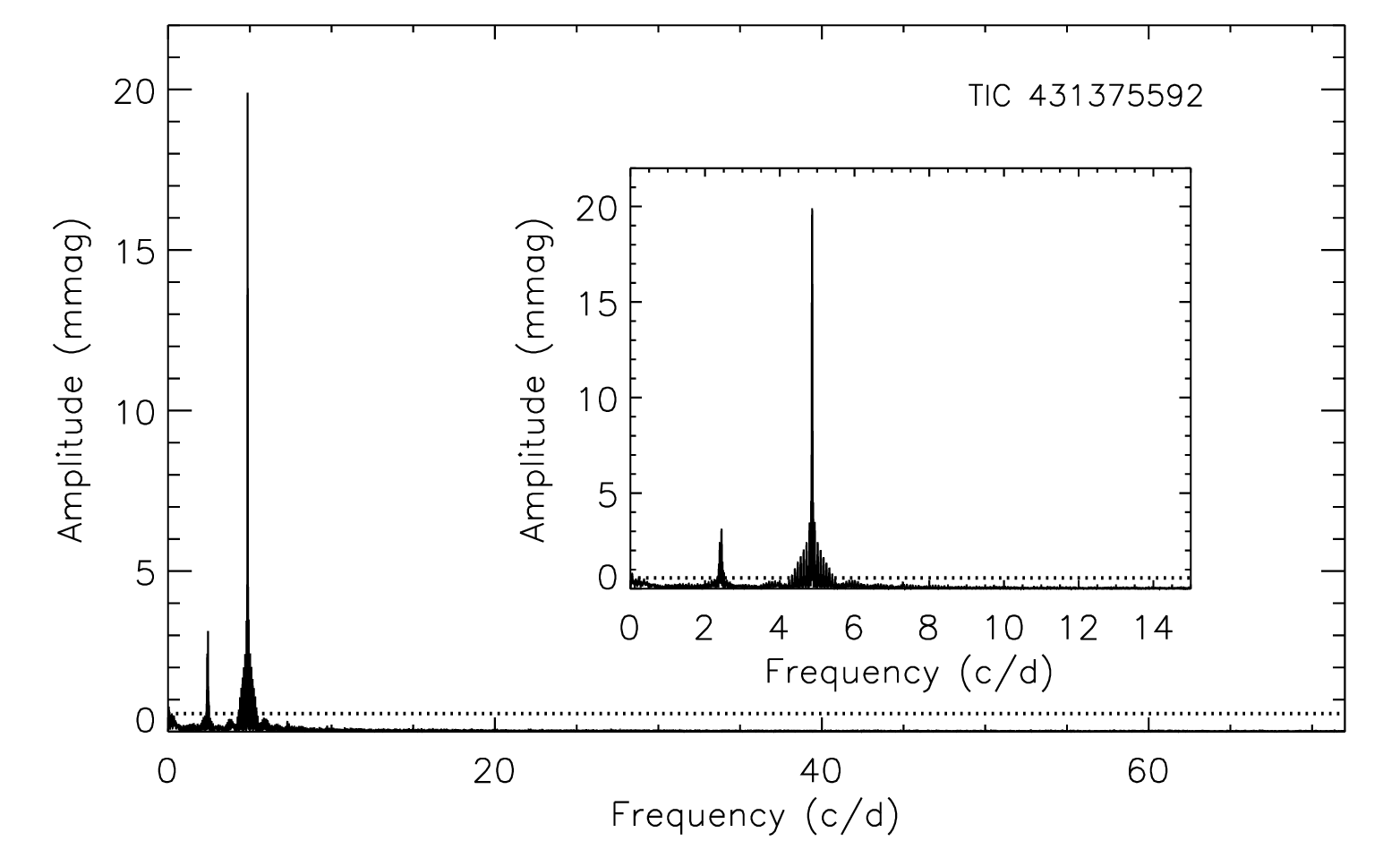}
 \end{minipage}
   \begin{minipage}[b]{0.24\textwidth}
  \includegraphics[height=3.5cm, width=1\textwidth]{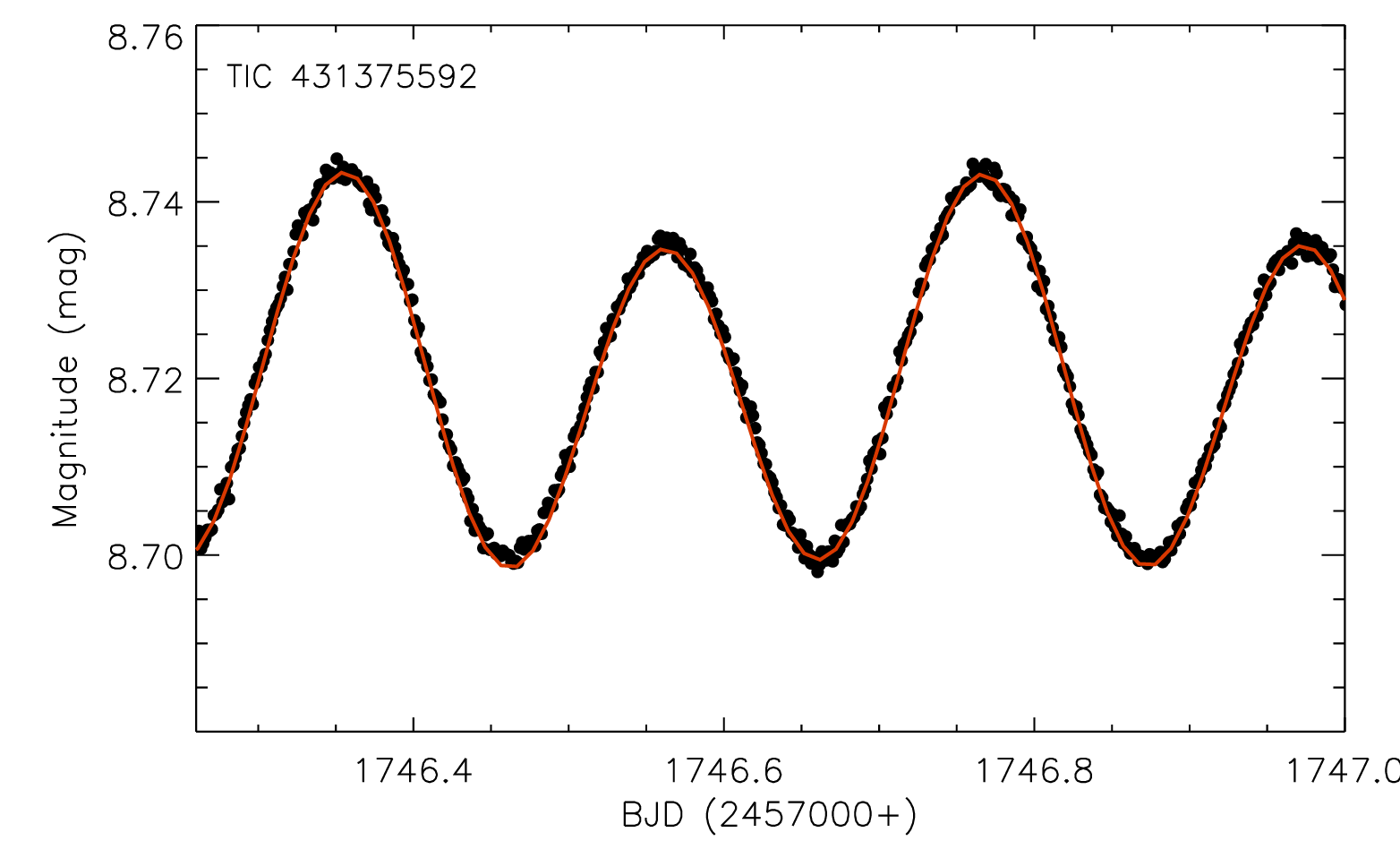}
 \end{minipage}
   \caption{Fourier spectra of the target stars and the theoretical frequency fits (solid red lines) to the observations (black dots). Dotted lines in the Fourier spectra represent the 4.5-$\sigma$ level.}
    \label{fourier}
\end{figure*}

As a result of the frequency analysis, we derived the pulsation frequencies of the targets and the list of the first ten frequencies and their S/N ratios are given in Table\,\ref{fre}. In this table, harmonic and combination frequencies are not listed. The full list of the frequencies is given in an electronic form. The Fourier spectra of the target stars and the theoretical frequency fits to the observations are shown in Fig.\,\ref{fourier}.

\section{DISCUSSION and CONCLUSIONS} 
\label{conc}

In this study, we present the analysis of some TESS field A-F type stars and revealed their pulsational characteristics. We selected the systems from the TESS field with inspection by eye and considering the some criteria such as $T_{\rm eff}$, frequency range. 

A frequency analysis was performed and all significant frequencies of the targets were derived. To classify the pulsation type of the system, we also calculated the pulsation constant and examine their position in the Hertzsprung-Russell (H-R) diagram. To do this, some important parameters of the targets were obtained. The extinction coefficient (A$_{\upsilon}$) was determined by using the galactic extinction map \citep{2005AJ....130..659A} with the help of the galactic coordinates and the Gaia eDR3 parallaxes \citep{2021A&A...649A...6G} of the targets. The calculated coefficients were then used to derive absolute magnitude ($M_{V}$), luminosity ($L$) and also bolometric magnitude ($M_{bol}$) with the same way described in the study of \cite{2021PASP..133h4201P}. The calculated parameters are given in Table\,\ref{para}.  

\begin{table*}
\begin{center}
\caption[]{Derived parameters for the targets and classification.}
\label{para}
 \begin{tabular}{llcccccc}
  \hline\noalign{\smallskip}
  &TIC    & A$_{\upsilon}$     &$M_{V}$  &  $M_{bol}$ & log ($L$/$L_\odot$) & Q     & Classification\\
  &number &(mag)\,$\pm$\,0.002 &(mag)    &(mag)       &                     & range & \\
   \hline\noalign{\smallskip}
1 &25537276  &0.08476&0.906\,$\pm$\,0.013 &0.987\,$\pm$\,0.014   &1.562 \,$\pm$\,0.033 & 0.011\,$-$\,0.025 & $\delta$ Scuti\\ 
2 &177422294 &0.09086&0.615\,$\pm$\,0.013&0.690\,$\pm$\,0.014   &1.678 \,$\pm$\,0.033  & 0.017\,$-$\,0.037 & $\delta$ Scuti\\
3 &252554307 &0.04787&2.149\,$\pm$\,0.011&2.229\,$\pm$\,0,012   &1.065 \,$\pm$\,0.031  & 0.021\,$-$\,0.056 & $\delta$ Scuti\\
4 &279874050 &0.04848&1.713\,$\pm$\,0.026&1.794\,$\pm$\,0,027   &1.239 \,$\pm$\, 0.046 & 0.028\,$-$\,0.031 & $\delta$ Scuti \\
5 &308447073 &0.03920&2.631\,$\pm$\,0.022&2.690\,$\pm$\,0.023   &0.872 \,$\pm$\,0.042  & 0.192\,$-$\,0.212 & $\gamma$ Doradus \\
6 &367910480 &0.00000&1.301\,$\pm$\,0.010&1.273\,$\pm$\,0.011   &1.403 \,$\pm$\,0.030  & 0.021\,$-$\,0.161 & Hybrid \\
7 &370599803 &0.06602&1.463\,$\pm$\,0.013&1.530\,$\pm$\,0.014   &1.339 \,$\pm$\,0.033  & 0.015\,$-$\,0.029 & $\delta$ Scuti\\
8 &395520454 &0.06427&2.539\,$\pm$\,0.025&2.608\,$\pm$\,0.026   &0.909 \,$\pm$\,0.045  & 0.050\,$-$\,0.120 & Hybrid \\
9 &400502366 &0.16216&0.276\,$\pm$\,0.014&0.273\,$\pm$\,0.015   &1.814 \,$\pm$\,0.034  & 0.032\,$-$\,0.465 & Hybrid \\
10&431375592 &0.03094&3.189\,$\pm$\,0.011&3.255\,$\pm$\,0.012   &0.648 \,$\pm$\,0.031  & 0.103\,$-$\,0.211 & Hybrid \\  
\noalign{\smallskip}\hline
\end{tabular}
\label{tablo1}
\end{center}  
\end{table*}

The resulting parameters were used to obtain the pulsation constant. Pulsation constant (Q) is defined by the following formula:\\

\begin{math}
\centering
Q=P\sqrt[]{\frac{\overline\rho}{\overline\rho{_\odot}}}
\end{math}\\

This equation can also be given as described by \cite{1990DSSN....2...13B}, \\

\begin{math}
\log(Q/P)=0.5\log\,g+0.1M_{bol}+\log\,T_{\rm eff}-6.456
\end{math}\\

\begin{figure}
   \centering
   \includegraphics[width=8cm, angle=0]{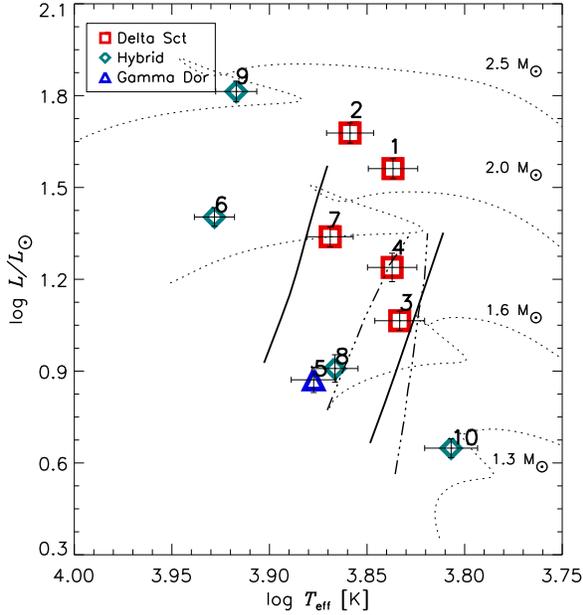}
   \caption{Position of the targets in the H-R diagram. The numbers are the same with Table\,\ref{fre} and show the stars. The solid and dot-dash lines represent the instability strips of $\delta$\,Sct and $\gamma$\,Dor stars respectively \citep{2005A&A...435..927D}. The dot lines are the evolutionary tracks taken from \cite{2016MNRAS.458.2307K}.}
   \label{hr}
   \end{figure}

In our study, the Q values of the target stars were calculated using the above equation and the parameters given in Table\,\ref{first} and Table\,\ref{para}. For some stars there are no $\log g$ parameters. For these stars we assumed the $\log g$ as 4.0. The range of the calculated Q values for each star is given in Table\,\ref{para}. By taking into account the frequency spectra and the Q values of the targets we made final pulsation classification. The Q constant is different for the $\delta$\,Sct and $\gamma$\,Dor stars. In the study of \cite{2002MNRAS.333..251H}, they presented the Q distributions for both pulsating star types. For $\delta$\,Sct and $\gamma$\,Dor stars, the Q is in the range of $\sim$0.008\,$-$\,0.063 and 0.200\,$-$\,1.260, respectively. Considering these Q values and the frequency spectra we classified the stars. The final classification is given in the Table\,\ref{para}. As a result of these investigation, we determined five $\delta$\,Sct, one $\gamma$\,Dor and four hybrid stars. The position of stars are shown in the H-R diagram as well. As can be seen from Fig.\,\ref{hr}, the $\delta$\,Sct star are located inside their instability strip and also $\gamma$\,Dor variable is placed very close its own domain. However, the hybrid stars are mostly located outside of both instability strips. This result is excepted for these variable stars according to the result found by \cite{2011A&A...534A.125U}. There are a few hypnotises to explain the existence of these variables such as wrong $T_{\rm eff}$, fast rotation and binarity \citep{2018BSRSL..87..137L, 2020svos.conf..353L}. No certain explanation has been found yet. Additionally, there is one target TIC\,431375592 which was announced as a possible variable star in the study of \cite{2002PASP..114..974B}. In our study, we confirmed the variability type of this systems. As a result of this study, we revealed ten A-F type pulsating stars which was not known or confirmed in the literature before. 

\newpage
\begin{acknowledgments}
This work has been supported in part by the Scientific and Technological Research Council (TUBITAK) under the grant number 120F330. The TESS data presented in this paper were obtained from the Mikulski Archive for Space Telescopes (MAST). Funding for the TESS mission is provided by the NASA Explorer Program. This work has made use of data from the European Space Agency (ESA) mission Gaia (http://www.cosmos.esa.int/gaia), processed by the Gaia Data Processing and Analysis Consortium (DPAC, http://www.cosmos.esa.int/web/gaia/dpac/\\consortium). Funding for the DPAC has been provided by national institutions, in particular the institutions participating in the Gaia Multilateral Agreement. This research has made use of the SIMBAD data base, operated at CDS, Strasbourq, France. 
\end{acknowledgments}

\section*{FUNDING}
This work has been supported in part by the Scientific and Technological Research Council (TUBITAK) under the grant number 120F330.

\section*{CONFLICT OF INTEREST}
The authors declare no conflicts of interest.



 \appendix

\setcounter{table}{2}
\begin{table*}[htb]
 \begin{scriptsize}
\begin{center}
\caption[]{Results of the frequency analysis. Harmonic and combination frequencies are not given in this list. The full list of the frequencies is given in electronic form.}
\label{fre}
 \begin{tabular}{llccc|ccccc}
  \midrule
  \hline\noalign{\smallskip}
   & \multicolumn{4}{c}{\hrulefill TIC\,25537276 \hrulefill} & \multicolumn{5}{c}{\hrulefill TIC\,177422294\hrulefill}  \\  
   & Frequency   &  Amplitude & Phase &S/N  &  &Frequency   &  Amplitude & Phase &S/N     \\
   & (d$^{-1}$)  & (mmag)     & (rad) &     &  &(d$^{-1}$)  & (mmag)     & (rad) &        \\
   \hline\noalign{\smallskip}
$\nu_1$ &7.55054\,$\pm$\,0.00002 &8.872\,$\pm$\,0.008 &0.2364\,$\pm$\,0.0001  & 534.6  & $\nu_1$ &8.30240\,$\pm$\,0.00016 &6.863\,$\pm$\,0.006&0.0739\,$\pm$\,0.0011  & 14 \\
$\nu_2$ &8.03904\,$\pm$\,0.00008 &2.237\,$\pm$\,0.008 &0.6364\,$\pm$\,0.0006  & 151.0  & $\nu_2$ &7.66622\,$\pm$\,0.00036 &3.318\,$\pm$\,0.006 &0.0640\,$\pm$\,0.0026 & 6.6 \\
$\nu_3$ &7.37013\,$\pm$\,0.00011 &1.739\,$\pm$\,0.008 &0.3471\,$\pm$\,0.0007  & 107.3  & $\nu_3$ &7.07205\,$\pm$\,0.00039 &3.208\,$\pm$\,0.006 &0.5485\,$\pm$\,0.0028  & 6.1 \\
$\nu_4$ &5.78908\,$\pm$\,0.00012 &1.506\,$\pm$\,0.008 &0.6576\,$\pm$\,0.0008  & 100.5  & $\nu_4$ &8.204374\,$\pm$\,0.00028 &4.808\,$\pm$\,0.006 &0.918\,$\pm$\,0.002  & 9.9 \\
$\nu_5$ &7.04380\,$\pm$\,0.00026 &0.706\,$\pm$\,0.008 &0.1234\,$\pm$\,0.0018  & 43.8  & $\nu_5$ &8.23038\,$\pm$\,0.00015  &2.438\,$\pm$\,0.006 &0.033\,$\pm$\,0.001 & 8.7 \\
$\nu_6$ &7.57891\,$\pm$\,0.00062 &0.321\,$\pm$\,0.008 &0.0344\,$\pm$\,0.0044  & 19.2  & $\nu_6$ &8.12435\,$\pm$\,0.00081  &2.438\,$\pm$\,0.006 &0.0852\,$\pm$\,0.0064  & 5.0 \\
$\nu_7$ &15.0991\,$\pm$\,0.00110 &0.168\,$\pm$\,0.008 &0.8044\,$\pm$\,0.0078  & 14.3  & $\nu_7$ &4.51330\,$\pm$\,0.00048  &2.551\,$\pm$\,0.006 &0.8588\,$\pm$\,0.0034  & 5.3 \\
$\nu_8$ &7.65999\,$\pm$\,0.00474 &0.103\,$\pm$\,0.008 &0.9871\,$\pm$\,0.0337  & 6.3  & $\nu_8$ &7.70823\,$\pm$\,0.00042  &2.723\,$\pm$\,0.006 &0.5254\,$\pm$\,0.0031  & 5.4 \\
$\nu_9$ &7.94174\,$\pm$\,0.00177 &0.111\,$\pm$\,0.008 &0.4499\,$\pm$\,0.0126  & 7.3  & $\nu_9$ &8.39643\,$\pm$\,0.000224  &2.929\,$\pm$\,0.006 &0.3438\,$\pm$\,0.0016  & 6.4 \\
$\nu_{10}$ &8.00864\,$\pm$\,0.00554 &0.101\,$\pm$\,0.008 &0.4312\,$\pm$\,0.0395  & 6.6  & $\nu_{10}$ &5.23151\,$\pm$\,0.00056  &2.272\,$\pm$\,0.006 &0.9086\,$\pm$\,0.0041  & 5.1 \\
  
\hline\noalign{\smallskip}
   & \multicolumn{4}{c}{\hrulefill TIC\,252554307\hrulefill} & \multicolumn{5}{c}{\hrulefill TIC\,279874050\hrulefill}  \\  
   & Frequency   &  Amplitude & Phase &S/N  &  &Frequency   &  Amplitude & Phase &S/N     \\
   & (d$^{-1}$)  & (mmag)     & (rad) &     &  &(d$^{-1}$)  & (mmag)     & (rad) &        \\
   \hline\noalign{\smallskip}
$\nu_1$ &7.57891\,$\pm$\,0.00003 &5.213\,$\pm$\,0.008 &0.3055\,$\pm$\,0.0002   & 288.6 & $\nu_1$ &12.85860\,$\pm$\,0.00032 & 9.709\,$\pm$\,0.001 &0.3965\,$\pm$\,0.1035 & 62.6 \\
$\nu_2$ &7.51810\,$\pm$\,0.00003 &5.651\,$\pm$\,0.008 &0.2651\,$\pm$\,0.0002   & 313.5 & $\nu_2$ &11.55165\,$\pm$\,0.00003 & 4.149\,$\pm$\,0.005 & 0.0212\,$\pm$\,0.0024  & 29.7 \\
$\nu_3$ &8.07147\,$\pm$\,0.00015 &1.321\,$\pm$\,0.008 &0.6219\,$\pm$\,0.00107  & 77.4 & $\nu_3$&12.86759\,$\pm$\,0.00016 & 5.091 \,$\pm$\,0.001 & 0.3965\,$\pm$\,0.0081  & 32.8 \\
$\nu_4$ &7.45324\,$\pm$\,0.00010 &1.748\,$\pm$\,0.008 &0.8083\,$\pm$\,0.0007  & 99.1 & $\nu_4$ &12.69569\,$\pm$\,0.00026 & 2.476 \,$\pm$\,0.001 & 0.0211\,$\pm$\,0.0081   & 28.4 \\
$\nu_5$ &7.39851\,$\pm$\,0.00012 &1.255\,$\pm$\,0.006 &0.5140\,$\pm$\,0.0025 & 70.6  & $\nu_5$ &12.89069\,$\pm$\,0.00015 & 2.438\,$\pm$\,0.006 & 0.2630\,$\pm$\,0.0025   & 14.7 \\
$\nu_6$ &8.00458\,$\pm$\,0.00053 &1.532\,$\pm$\,0.008 & 0.1942\,$\pm$\,0.2702 &  89.8 & $\nu_6$ &12.90520\,$\pm$\,0.00039 & 2.375\,$\pm$\,0.006  & 0.2702 \,$\pm$\,0.0046   & 15.4 \\
$\nu_7$ &5.76070\,$\pm$\,0.00018 &0.953\,$\pm$\,0.008 &0.5921\,$\pm$\,0.0013 & 54.9 & $\nu_7$ &12.84552\,$\pm$\,0.00054 & 1.608\,$\pm$\,0.004  & 0.7985 \,$\pm$\,0.0052  & 10.1 \\
$\nu_8$ &7.68634\,$\pm$\,0.00015 &0.122\,$\pm$\,0.008 &0.2223\,$\pm$\,0.0010 & 66.8 & $\nu_8$ &11.58353\,$\pm$\,0.00065 & 1.327\,$\pm$\,0.0042 & 0.7832 \,$\pm$\,0.0054  & 9.2 \\
$\nu_9$ &7.00933\,$\pm$\,0.00066 & 0.262\,$\pm$\,0.008 &0.72455\,$\pm$\,0.0047 & 14.9  \\
$\nu_{10}$ &7.79377\,$\pm$\,0.00054 &0.144\,$\pm$\,0.008 & 0.1784 \,$\pm$\,0.0038 & 8.16 \\
\hline\noalign{\smallskip}
\end{tabular}
\label{tablo1}
\end{center} 
 \end{scriptsize}
\end{table*}

\setcounter{table}{2}
\begin{table*}[htb]
 \begin{scriptsize}
\begin{center}
\caption[]{Continuation.}
\label{fre}
 \begin{tabular}{llccc|ccccc}
  \midrule
  \hline\noalign{\smallskip}
   & \multicolumn{4}{c}{\hrulefill TIC\,308447073\hrulefill} & \multicolumn{5}{c}{\hrulefill TIC\,367910480\hrulefill}  \\  
   & Frequency   &  Amplitude & Phase &S/N  &  &Frequency   &  Amplitude & Phase &S/N     \\
   & (d$^{-1}$)  & (mmag)     & (rad) &     &  &(d$^{-1}$)  & (mmag)     & (rad) &        \\
   \hline\noalign{\smallskip}
$\nu_1$ & 3.39196\,$\pm$\,0.00012 &1.820\,$\pm$\,0.008 &0.9407\,$\pm$\,0.0007  & 104.8 & $\nu_1$ &15.6652\,$\pm$\,0.00003 & 0.516\,$\pm$\,0.008 &0.5253\,$\pm$\,0.0025 & 22.8 \\
$\nu_2$ &3.57380\,$\pm$\,0.00040 &5.359\,$\pm$\,0.008 &0.7289\,$\pm$\,0.0026   & 30.7 & $\nu_2$ &4.04676\,$\pm$\,0.00002 & 0.464\,$\pm$\,0.006 & 0.5525\,$\pm$\,0.0023  & 8.5 \\
$\nu_3$ &3.71073\,$\pm$\,0.00054 &0.358\,$\pm$\,0.008 &0.1172 \,$\pm$\,0.0035  & 21.3 & $\nu_3$&6.10102\,$\pm$\,0.00004 & 0.438 \,$\pm$\,0.007 & 0.3323\,$\pm$\,0.0034  & 10.9 \\
$\nu_4$ &3.43462\,$\pm$\,0.00061 &0.380\,$\pm$\,0.008 & 0.5469 \,$\pm$\,0.0039 & 21.4 & $\nu_4$ &19.87434\,$\pm$\,0.00004 & 0.379\,$\pm$\,0.007 & 0.4481\,$\pm$\,0.0028   & 21.7 \\
$\nu_5$ &3.35380\,$\pm$\,0.00069 &0.318\,$\pm$\,0.006 &0.2343 \,$\pm$\,0.0044  & 18.1  & $\nu_5$ &3.64199\,$\pm$\,0.00004 & 0.340$\pm$\,0.006 & 0.8268\,$\pm$\,0.0031   & 6.3 \\
        &                        &                    &                        &       & $\nu_6$ &2.59348\,$\pm$\,0.00004 &  0.298$\pm$\,0.006 & 0.3935\,$\pm$\,0.0035  & 6.1  \\
        &&& & & $\nu_7$ &3.34585\,$\pm$\,0.00004 & 0.285$\pm$\,0.006 & 0.0536\,$\pm$\,0.0037   & 5.1 \\
        &                        &                    &                        &       &  $\nu_8$ &3.54846\,$\pm$\,0.00003 & 0.275$\pm$\,0.006 & 0.0332\,$\pm$\,0.0039   & 4.9   \\
        &                        &                    &                        &       & $\nu_9$ &5.76989\,$\pm$\,0.00005 & 0.245$\pm$\,0.006 & 0.3817\,$\pm$\,0.0043   & 5.3    \\
\hline\noalign{\smallskip}
   & \multicolumn{4}{c}{\hrulefill TIC\,370599803\hrulefill} & \multicolumn{5}{c}{\hrulefill TIC\,395520454\hrulefill}  \\  
   & Frequency   &  Amplitude & Phase &S/N  &  &Frequency   &  Amplitude & Phase &S/N     \\
   & (d$^{-1}$)  & (mmag)     & (rad) &     &  &(d$^{-1}$)  & (mmag)     & (rad) &        \\
   \hline\noalign{\smallskip}
$\nu_1$ &18.32261\,$\pm$\,0.00009 &3.664\,$\pm$\,0.001 &0.9809 \,$\pm$\,0.0006  & 84.5 & $\nu_1$ &2.82490\,$\pm$\,0.00052 & 1.924\,$\pm$\,0.006 &0.5297\,$\pm$\,0.0009 & 44.1 \\ $\nu_2$ &9.19831\,$\pm$\,0.00010 &3.507\,$\pm$\,0.001 & 0.5446\,$\pm$\,0.0007  & 51.8 & 
$\nu_2$ &2.76637\,$\pm$\,0.00118 & 1.870\,$\pm$\,0.003 &0.9027\,$\pm$\,0.0009 & 42.5 \\
$\nu_3$ &9.80307\,$\pm$\,0.00017 &2.270\,$\pm$\,0.001 & 0.6731\,$\pm$\,0.0011  & 36.1 & $\nu_3$ &3.06876\,$\pm$\,0.00016 & 1.363\,$\pm$\,0.001  & 0.0541\,$\pm$\,0.0012  & 31.7 \\
$\nu_4$ &9.86653\,$\pm$\,0.00020 &1.948\,$\pm$\,0.001 & 0.1550\,$\pm$\,0.0013  & 30.5 & $\nu_4$ &2.56933\,$\pm$\,0.00015 & 0.261\,$\pm$\,0.006 & 0.8112\,$\pm$\,0.0025   & 5.9 \\
$\nu_5$ &12.63869\,$\pm$\,0.00023 &1.509\,$\pm$\,0.001 &0.35997 \,$\pm$\,0.0016 & 23.5 & $\nu_5$ & 3.38871\,$\pm$\,0.00092 & 0.219\,$\pm$\,0.001  & 0.398\,$\pm$\,0.0068   & 5.1 \\
$\nu_6$ &8.72465\,$\pm$\,0.00028 &1.256\,$\pm$\,0.001 & 0.9525\,$\pm$\,0.0019 & 23.6 & $\nu_6$ &2.98097\,$\pm$\,0.00102& 0.219\,$\pm$\,0.001  & 0.0258\,$\pm$\,0.0072  & 5.5 \\
$\nu_7$ &10.05893\,$\pm$\,0.00029 &1.207\,$\pm$\,0.001 & 0.3085\,$\pm$\,0.0020 & 18.6 & $\nu_7$ &6.14338\,$\pm$\,0.00132& 0.186\,$\pm$\,0.001 & 0.0336\,$\pm$\,0.0093  & 6.8\\
$\nu_8$ &10.92167\,$\pm$\,0.00029 &1.120\,$\pm$\,0.001 &0.0247\,$\pm$\,0.0020 & 25.7 \\
$\nu_9$ &13.64733\,$\pm$\,0.00066 & 0.873\,$\pm$\,0.001 & 0.0247\,$\pm$\,0.0028 & 13.7  \\
$\nu_{10}$ &11.41436\,$\pm$\,0.00042 & 0.826\,$\pm$\,0.001 & 0.3125 \,$\pm$\,0.0029 & 16.0 \\
   \hline\noalign{\smallskip}
\end{tabular}
\label{tablo1}
\end{center} 
 \end{scriptsize}
\end{table*}

\setcounter{table}{2}
\begin{table*}[htb]
 \begin{scriptsize}
\begin{center}
\caption[]{Continuation.}
\label{fre}
 \begin{tabular}{llccc|ccccc}
  \midrule
  \hline\noalign{\smallskip}
   & \multicolumn{4}{c}{\hrulefill TIC\,400502454\hrulefill} & \multicolumn{5}{c}{\hrulefill TIC\,431375592\hrulefill} \\  
   & Frequency   &  Amplitude & Phase &S/N  &  &Frequency   &  Amplitude & Phase &S/N     \\
   & (d$^{-1}$)  & (mmag)     & (rad) &     &  &(d$^{-1}$)  & (mmag)     & (rad) &        \\
   \hline\noalign{\smallskip}
$\nu_1$ &6.59953\,$\pm$\,0.00007 &3.398\,$\pm$\,0.001 & 0.4157\,$\pm$\,0.0005  & 112.2  & $\nu_1$ &4.86512\,$\pm$\,0.00005 &19.915\,$\pm$\,0.009 & 0.0034\,$\pm$\,0.0007  & 756.6 \\
$\nu_2$ &5.93556\,$\pm$\,0.000132 &2.003\,$\pm$\,0.001 &0.8206\,$\pm$\,0.0009  & 57.9  & $\nu_2$ &2.43893\,$\pm$\,0.00006 &2.912\,$\pm$\,0.009  & 0.0833\,$\pm$\,0.0002 & 56.2 \\
$\nu_3$ &3.71104\,$\pm$\,0.00022 &1.192\,$\pm$\,0.001 &0.0247\,$\pm$\,0.0015  & 26.8  & $\nu_3$ &2.39483\,$\pm$\,0.00024 &1.902\,$\pm$\,0.009  & 0.9105\,$\pm$\,0.0003  & 36.3 \\
$\nu_4$ &7.95496\,$\pm$\,0.00022 &1.157\,$\pm$\,0.001 &0.3522\,$\pm$\,0.0015  & 43.3  & $\nu_4$ &2.37720\,$\pm$\,0.00043 &0.907\,$\pm$\,0.009  &  0.56145\,$\pm$\,0.0006  & 17.2 \\ 
$\nu_5$ &1.35331\,$\pm$\,0.00044 &0.598\,$\pm$\,0.001 &0.4110\,$\pm$\,0.0030  & 10.1  \\ 
$\nu_6$ &9.64449\,$\pm$\,0.00045 &0.581\,$\pm$\,0.001 &0.4577\,$\pm$\,0.0031  & 21.6  \\ 
$\nu_7$ &4.68162\,$\pm$\,0.00056 &0.468\,$\pm$\,0.001 &0.6521\,$\pm$\,0.0038  & 10.8  \\ 
\hline\noalign{\smallskip}
\end{tabular}
\label{tablo1}
\end{center} 
 \end{scriptsize}
\end{table*}

\end{document}